\documentclass[11pt]{report}
\usepackage{graphicx}
\usepackage{framed}
\usepackage{listings}
\usepackage{amsthm}
\usepackage{textcomp,    % for \textlangle and \textrangle macros
            xspace}
\newcommand\la{\bf\textlangle\xspace}  % set up short-form macros
\newcommand\ra{\bf\textrangle\xspace}
\newtheorem*{mydef}{Definition}
\usepackage{amsfonts}
\usepackage{a4}
\usepackage{bnf}
\usepackage[top=1in, bottom=1in, left=1in, right=1in]{geometry}\author{A.Shafarenko}
\date{}
\def\ak{{\textsf{A\kern-1.5pts\kern-1ptt\kern-1ptr\kern-1.8pta}}\kern-2pt{\it K\kern-2ptahn}}
\lstnewenvironment{code}[1][]%
{
   \noindent
   \minipage{\linewidth}
   \vspace{0.5\baselineskip}
   \lstset{basicstyle=\ttfamily\footnotesize,frame=single,#1}}
{\endminipage}
\begin{document}
\begin{titlepage}
\begin{center}
\vspace{0.5in}
{\Huge AstraKahn: A Coordination Language for Streaming Networks}\\
\vspace{0.5in}
{\Huge\it Language Report v0.13 \\
\today\\}
\vspace{0.5in}
{\Large
A.Shafarenko\\
\ \\
Compiler Technology and Computer Architecture Group\\
University of Hertfordshire
}
\end{center}
\begin{abstract}
This is a preliminary version of the language report. It contains key definitions, specifications and some examples, but lacks completeness. The full document will include Chapter 3 (Data and Instrumentation Layer) and an appendix giving the complete syntax and some whole program examples. The purpose of the present document is to fix the concepts and major features of the language and to enable the production of the definition document that is required for implementation.

{\bf Acknowledgements} We acknowledge the help of  colleagues in shaping up the present report. The subject matter was repeatedly discussed with Dr Raphael Poss to whom we are indebted for drawing our attention to map-reduce parallelism, and who also provided  invaluable critique elsewhere. We also wish to thank Dr Raimund Kirner for his input in regard of reat-time considerations and possible integration of the language with a real-time programming environment, and Mr Pavels Zaichenkovs for suggestions that improved the the syntax of MDL.
\end{abstract}
\end{titlepage}
\chapter{The Topology and Progress Layer}
\section{Motivation}

\ak\ is based on Gilles Kahn's model of process network (KPN). The program is represented as a graph whose edges are
streams of messages. In KPN the vertices of the graph are pure prefix-monotonic stream-to-stream functions. In \ak\ this
is also the case except that a special type of vertex, called synchroniser, can also be nondeterministic, in which case it can be
assumed that there exists one more stream in the system, an {\em oracle} stream, which is fed to all synchronisers as an additional
edge. With this (assumed) edge  included, and also assuming that any nondeterministic choices that a synchroniser makes are functionally
dependent on the value of the current oracle message, the synchroniser is also a pure function of its input streams.
The output of a deterministic program should (provably) be independent of the values of
the oracle stream. \footnote{In the case of a significantly nondeterministic program, for each output stream a projection
function should be defined whose output is independent of the oracle values. The network thus augmented can still be amenable
to ordinary correctness proofs} All types of vertex in \ak\ are prefix-monotonic.

The intention of \ak\ is thus to refine and structure KPNs. The purpose of the structuring and the refining is dual. Firstly and
most importantly, a KPN is a theoretical model which has nice properties but those are only available under an interpretation of
the model that does not limit the resources. In particular, the KPN streams have an indefinite buffering capacity, and consequently
the progress of each vertex is dependent solely on the availability of its input data. Here we assume the standard interpretation of
a vertex as a process that computes and sends out prefixes of the output streams as soon as a sufficient prefix of each of the input streams
becomes available.  In reality, all sorts of resources are required for the evaluation of a KPN, and those come from a limited
pool and have to be shared amongst the vertices and their connecting streams. The issue of progress becomes quite complicated
and can only be addressed by an elaborate regulatory mechanisms not found in the original model. \ak\ attempts to offer
such mechanisms by using a behavioural classification of the vertices and a self-regulatory network composition.

The second purpose of refinement comes from the standard software engineering agenda: abstraction, encapsulation and
hierarchical development. Those issues require components and component interfaces. Here \ak\ attempts to provide what
is known as a {\em coordination} language, i.e. a language in which the components' connection, communication and synchronisation
can be expressed. It is thus a form of software glue for the components it coordinates. As a result, the components need not,
and in most cases should not attempt to, address concurrency; they can limit themselves to either sequential or data-parallel
execution. Moreover, with the exception of synchronisers, which are programmed in \ak, the rest of the vertices can be programmed
in any (concurrency oblivious)  language. \ak\ is therefore a {\em coordination}, rather than a fully-fledged programming,
language.

\ak\ is best seen as a three-layer construction. The bottom, {\em Topology and Progress} Layer (TPL) defines the topology and provides the basic concurrency
mechanism over it, based on the concept of communication pressure, i.e. a consumer's denial or demand of the producer's output.
Pressure is a key regulatory mechanism of \ak\ and is present
in the behavioural definitions of all vertices and streams. At the bottom layer, the network is represented as a set of named
vertices connected by named channels, but the channels stream data-agnostic messages. Vertices provide their data constraints
and relate their input to their output data properties in a separate, {\em Constraint Aggregation} Layer (CAL). This provides
for sufficient characterisation in order to ensure type safety and a seamless interface with a vertex programming language.
Finally, the \ak\ approach to vertex refinement ensures that only synchronisers possess a persistent state, while the rest
of the vertices are stateless. Consequently, it is \ak\ that must provide data storage and communication facilities, which
depend on the nature of the data (available from the CAL) and the concurrency and distribution strategies that come
from the TPL. This is achieved in the {\em Data and Instrumentation} Layer, DIL,
where also monitoring of the data usage and movement is implemented.

\section{Philosophy}

First of all, the semantics of \ak\ on the TPL is described in terms of structures put in place for the {\em coordinator}, i.e. a controlling
agent, or indeed a group of agents, responsible for progress and communication of the KPN vertices. The vertices are
connected by stream-carrying {\em channels}  using {\em wiring patterns}. There are a two kinds of vertex: a box and a synchroniser. Boxes
are stateless functions that  read messages from one (or in some cases two)  input channels. Boxes read an input message, do some computation
and send messages to output channels. Synchronisers are vertices that can have any number of input channels. Synchronisers
have an internal state and generally accept messages from each input channel in some states, while in any of the other states
the channel is {\em blocked} until a state transition brings the synchroniser to a state in which messages from the
channel are accepted, thus unblocking it. Synchronisers are able to store received messages and retrieve them for the sole purpose
of sending them on, either as they are, or combined with other stored or received messages and with trivial message extensions
computed by the synch itself. The state transitions of a synchroniser can depend on the content of the current message but never on
that of a stored one. In other words, a synchroniser is a finite state machine for joining
messages and sending them on to its output channels. It should be noted that {\em all} vertices
are pure {\em stream-to-stream} prefix-monotonic functions that map the totality of their input streams
to the totality of their output streams; the differentiation between boxes and synchronisers is based on their response to {\em individual}
messages: boxes are "context-free" in that their responses to individual messages do not depend on past histories (while nothing
can depend on the future in a prefix-monotonic vertex anyway), while in the case of synchronisers such a dependency may exist.

Boxes and synchronisers are connected by {\em segmented} channels. A channel carries a stream that consists of messages and
possibly segmentation marks. The latter can be thought of as ``brackets''. The intention of the brackets is to mark the beginning
and the end of a sequence. Boxes can take such a sequence into a single result-message, or produce a sequence from a single message,
or even ignore the segmentation structure and respond to each input message individually. The box code does not see the segmentation marks;
instead it is declared as having a certain type of bracketing behaviour so that the coordinator can take care of the brackets.
A synchroniser sees segmentation marks as data and makes transitions based on them according to its transition graph. Sequences
may in turn consist of sequences in their own right, hence a channel can have a {\em bracketing depth} (or  depth for short) greater than one.
However, each message on a given channel will under all circumstances find itself between the same number of brackets, hence
the depth is static, and is a characteristic of a channel available from the TPL\footnote{this is in sharp contrast with the characteristics of messages,
such as type, which are only available from the CAL}. Boxes responding to a sequence appear to contradict the earlier statement about
statelessness; but in fact the statement still holds. A box that accepts a sequence can be thought of as one reading two messages from
its input(s) and replacing one of them with an intermediate result to be read next time, until the whole sequence is replaced by the final result.
Each of these steps leaves the box completely state-free. In the opposite direction, a box that produces a sequence of messages from
a single message can be thought of as producing just two messages: one for the output and the other replacing the input message,
so that the next time the input message is read the production of the output sequence may continue until finally it is completed and
then no replacement for the input message takes place. Such behaviour is only virtual under the normal circumstances, but it becomes
physically manifest under pressure; we shall dwell on this later. We only wish to remark at this point that boxes that reduce a sequence
to a single message may also have declarable algebraic properties, such as commutativity and associativity with respect to the input stream.

Now to wiring patterns. They form a small set: only 5 in total,  describing connections between nodes in a hierarchical way. A pattern
is applied to its operand networks, either vertices or groups of nodes already wired up with some wiring patterns. The pattern identifies
input/output channels of the operand(s) with one another and with the input/output channels of the result. Four out of five patterns are finite,
applicable to one or two operands. The fifth one is infinite as it infinitely replicates the single operand network
and wires up an infinite chain. That last pattern can also be dynamic, in the sense that the wiring takes place step-by-step at run time, which is the most complex topological and regulatory behaviour available in TPL.

Synchronisers are defined in the
input language of \ak, while boxes are specified in a box programming language and are subject to a {\em box contract}. The contract sets out acceptable
behaviour for a box and is partly cooperative (i.e. cannot be inforced, but any guarantees that \ak\ makes are subject to the fulfilment of the contract on
behalf of all the boxes) and partly enforceable.  \ak\ accepts any box programming language that enables the programmer to fulfil the contract.
The interface between a box and the \ak\ run-time system for any valid implementation is defined by the \ak\ Box-API for each supported box language.

The TPL is the base layer of \ak, whcih means that it can be understood without reference to other layers, and it is also the case that the other layers
need the TPL as a basis, without which their effects cannot be defined or understood. Besides the topology of the network and the choice of the vertices,
the TPL sets out the basic regulatory structure of \ak, whereby the run-time system can determine which boxes should be run and at what moment
in order to guarantee (within certain assumptions) that any application that can produce its intended output and terminate will manage to do so
and that the concurrency of the boxes given the finite set of platform resources will not be unduly restricted. This is achieved by the concepts
of {\em positive and negative pressure}. In \ak\
positive pressure is the condition of a channel associated with reduced storage capacity in its FIFO queue. The pressure is quantified as the
number of queue elements held at any given time. Positive pressure can become {\em critical} at the point when the channel is unable to provide
a fresh FIFO element and becomes {\em blocked}. A blocked channel refuses
any further input from the producer end until its queue is shortened by a message delivery, that is, when the pressure
drops to a subcritical level; consequently the positive pressure directly affects the producer. By contrast,
negative pressure is a condition associated with the consumer end of an {\empty} channel. Negative pressure is created by the consumer at the point in
time when the consumer has a demand for more messages that is not immediately satisfiable. Such demand is quantified as the number of messages that would be accepted by the consumer {\em immediately}. The consumer under negative pressure is unable to process messages, just like the producer under supercritical positive pressure. Conversely, a positive pressure in the input channel and a negative pressure in an output one stimulate the node
to increase its processing rate. These simple considerations are fundamental to understanding the regulatory mechanism of \ak\ TPL.

Pressure propagates across the network due to the fact that the same node can be a producer for one neighbouring node and a consumer for another. The producer node is under positive pressure against producing too many messages too soon; if it does produce them, the channel becomes blocked,
but the producer is inhibited too, as it is unable to accept further {\em input} messages to itself even though it may have finished its current
computation (but obviously, not the communication of the results).
This can lead to the corresponding input channel building up pressure and eventually blocking its producer. As a result positive pressure
propagates backwards, which explains the equivalent term ``back pressure''. Negative pressure can be exerted by the environment consuming
the outputs of the whole network to indicate that the production rate is insufficient, but it can also be exerted
by synchronisers, when they have to block one of their input channel due to the unavailability of matching messages on another.
For example, this may occur when a synch performs a zip operation, in a situation when one of the two producers is
systematically slower than the other. The former is then put under a negative pressure corresponding to the positive pressure exerted on the latter.
Like positive pressure, negative pressure may become critical,  triggering a concurrency reallocation. For instance, a box under
a positive input pressure and a negative output pressure may be given more cores on a multicore platform to speed it up internally, or several
copies of the box may be applied to consecutive elements of the input queue at the same time, depending on its algebraic properties.
Also a chain of boxes under negative pressure at the destination end can propagate that pressure back to the input of the chain in the same way
as positive pressure back-propagates. Unlike positive pressure, however, the negative one does not {\em necessarily} lead to any
definite behavioural change; a TPL implementation is at liberty to ignore negative pressure without invalidating the semantics, but it is not {\em efficient}
to do so under normal circumstances. Positive pressure can only be ignored until it becomes critical. The critical positive pressure level
is not defined either, but due to the finiteness of the resources, any channel is bound to go critical at some point if the consumer stops
to consume.

The TPL defines:
\begin{enumerate}
\item classes of boxes, their algebraic properties, their effects on channel segmentation and the behaviour under both types of pressure.
\item classes of channels with respect to pressure conductance
\item the language for synchronisers, including:
	\begin{itemize}
		\item the structure of the state, the specification of the state transitions and the associated storage/retrieval behaviour
		\item pressure creation and conductance
		\item nondeterminism and fairness policies
	\end{itemize}
\item the static wiring patterns
\item subnetwork encapsulation facility
\end{enumerate}

In the sequel we will contribute a section on each of the above bullets.

\section{Boxes}

\subsection{Channel segmentation}

As mentioned above, \ak\ channels carry bracketed sequences of messages, each message finding itself on the same bracketing level.
It is easy to see that brackets in such a channel can only occur in certain combinations. To start with, let us assume that the
channel does not contain sequences or subsequences of zero messages and that the bracketing depth is positive, too: $d>0$.
Such a sequence must start with $d$ opening brackets and a message. Further into the sequence brackets can only
be found in the following combination:
\[
\underbrace{)\ldots)}_k \underbrace{(\ldots(}_k\,,
\]
where $k\le d$. We shall denote that combination as a segmentation mark $\sigma_k$. Finally at the end of the sequence, when no further
messages are to be communicated, we must expect $d$ closing brackets. Given that $d$ is statically known for every channel, the
brackets at both ends of the sequence need not be communicated. The end-of-stream mark is still required to detect stream termination,
but it could be one mark for all stream depths. For convenience we shall denote the end-of-stream mark as $\sigma_0$, since it cannot be
misinterpreted as an ordinary segmentation mark. It should be noted that segmentation marks (including $\sigma_0$, which we will also call
a segmentation mark for convenience) are messages in their own right and that they are distinct from any data messages
that travel along the channel.

The set of segmentation marks supports empty (sub-)sequences but only when they happen at the full channel depth. Such sequences
may conveniently be thought of as missing messages. They are easily detected by discovering two segmentation marks in the channel
following one directly after the other.

\subsection{Productors}

We are now ready to consider the first category of nodes in \ak\, the productors. A productor has one input channel and one or more output channels.
It responds to a message received on the input channel by producing some output, and then terminates, and is re-launched by the coordinator
to receive a new message. The productor is a box, hence its response does not depend on any previous input messages as it is a pure function
of one input message. Two classes of productor are distinguished based on the output as follows.

\paragraph{Transductors}
A transductor is a productor that responds with no more than one output message on each of its output channels. If the input message is a $\sigma_k$,
it is passed on to all the output channels of the box by the coordinator, bypassing the box code.
The transductor thus only sees data messages. If any of the output channels
is blocked then the input channel will be refused progress and the box will not be activated until the pressure on all its output channels becomes subcritical.
Under sufficient negative pressure on all the output channels and large enough positive pressure on the input channel, the box code can be applied
in parallel to several input messages. We call this form of run-time adaptation {\em proliferation}. Exact proliferation thresholds
depend on the physical platform and the \ak\ implementation, they can vary during the run time as the run-time system adapts
to the performance characteristics of the code, but the coordination programmer can, and is expected to, rely on proliferation
as a strategy for capturing concurrency in program design.

The transductor part of the Box-API contains functions that enable it to read the input message and form zero or one output message
per output channel. The messages are only made available to the output channels upon the box termination.

\paragraph{Inductor}
An inductor is a productor that responds to a single message from the input channel with a sequence of messages on each of its output channels.
Before the input stream is passed to the inductor, each $\sigma_k$ in it with $k>0$ is replaced by $\sigma_{k+1}$, and a $\sigma_1$ is inserted
between every two consecutive data messages.  The inductor, just like the transductor, does not see segmentation marks, they are bypassed from
the input to all the output channels by the coordinator when encountered at the input of the inductor.

The box is not run if any of the output channels is blocked, in which case pressure is passed back in the same manner as it is with the transductor. Otherwise
the box is launched, it reads its input message and produces at most one message per output channel. The set of messages is then passed on to the Box-API,
which returns a flag indicating whether or not the box can continue to compute sets of output messages. If so then the next set is prepared, again at most one message per output channel, and the new flag is obtained, etc., until the box finishes its work and terminates despite the permission to continue.

Alternatively, if the flag indicates that the box is not allowed to continue
(which, for example, is the case when one or more of the output channels are blocked) and the box still has work to do,
then it must prepare a continuation message. The continuation message is a valid input message that, if received by the box, will cause the production
of exactly the same output messages as the ones that the box would have produced had it been allowed to continue.
The continuation message (or {\it continuation}, for short) will then be accepted by the Box-API and will replace the current
input message to the inductor.

The obligation to produce a valid continuation is part of the inductor contract with the TPL. This seems to be a restriction on the box
algorithm since, in order to fulfil the contract, the box must not keep anything important for its algorithm outside its input/continuation messages, not only
initially (which is guaranteed by the statelessness of the box), but also after every act of output. In reality this is not a severe restriction.
It is always possible to expand the input message with additional items by passing it through a transductor that adds and initialises those items.
Then the inductor could use them to store the state of its computation between acts of output.

By now it should be clear what the pressure conductance mechanism of the inductor should look like. Under positive pressure on any of the output channels,
an inductor will be forced to produce a continuation which replaces the original input message on the input channel. Progress on the input channel is
denied until the output becomes acceptable and the input message in question is finally consumed. Importantly, the box cannot stall with state inside
under any circumstances.

Negative pressure applied to all the output channels of an inductor simultaneously and combined  with positive pressure on the input, can lead to
proliferation as well. The amount of speculation involved in establishing the proliferation threshold here would be greater since each instance of
the inductor produces arbitrarily large output in the general case. Under such circumstances, running a parallel copy of the inductor on a message following behind the current one on the input channel is fraught with jittery output: until the work of the first copy ends, the output rate
will not increase at all, and at the moment of termination it will leap up, since the second copy will have been active and will have accumulated
some (possibly large amount of) output. Similar discontinuities of the rate are possible with more than two
parallel activities; however with a very large number of them compared to the average production length of an individual productor the
jitter will have the tendency to average down.

It should also be noted that the coordinator has access to the input {\em and} every continuation record of the productor. In the CAL
the box may define where in those messages a clue can be found to determine the amount of work likely to be required to complete the
processing of one record. Based on that and any instrumentation feedback from the DIL, it could be possible to auto-tune the proliferation
threshold dynamically.

\subsection{Reductors}

Reductors are boxes that take a list of input messages into a single output message. As an example of a reductor consider a function
that adds up the numbers contained in a list of messages and that outputs a message carrying the total. In general a reductor requires
a list of values and an initial value to perform its computation:
\[
a\oplus b_1\oplus b_2\oplus \ldots b_n\,,
\]

where the left associative operation $a \oplus b$ is  defined by a function $\oplus: \tau_a\to\tau_b\to\tau_a$ for any $a\in\tau_a$ and $b\in\tau_b$.
The computation has to proceed left to right as defined, but even so the operator $\oplus$ can be order agnostic, meaning that

\[
a\oplus b_1\oplus b_2\oplus \ldots b_n=a\oplus b_{\pi_1}\oplus b_{\pi_2}\oplus \ldots b_{\pi_n}\,.
\]
for any permutation $\pi_k$ of the natural series $1,\ldots, n$. When that is the case we shall call the reductor  {\em dyadic unordered}, otherwise it is {\em dyadic ordered}.

An important particular case of a reductor is when $\tau_a=\tau_b=\tau$. The reductor of this kind only requires a single list and will be called {\em monadic}.
A monadic reductor can similarly be ordered or unordered, and also can be segmentable, when the operator $\oplus$ is known to be associative
(but not necessarily commutative).

There are consequently two classes of dyadic reductor and three of monadic reductor. We are now prepared to describe the behaviour of reductors in detail.

\paragraph{Dyadic reductors.} Two channels are used, one for $a$s, and the other for lists of $b$s. The box is run when the coordinator detects the presence of  an $a$ on the first input channel and a $b$ on the second. The segmentation marks on the first channel are used in the same
manner as those of an inductor in respect of the output channels 2 and up. The segmentation marks on the second channel are used as list delimiters for the $b$s; they are also passed through to the reduction output, i.e. channel 1: any $\sigma_k$ with $k>1$ is sent out as
$\sigma_{k-1}$ , while $\sigma_1$ is removed, and $\sigma_0$ is passed on unaltered. These rules ensure that the second input channel can carry a sequence of depth two or more and that the reductor is applied to the innermost sequence and replaces it, along with its enclosing brackets, by the result of the computation. This way the dyadic reductor can be applied repeatedly, or different dyadic reductors can be applied one after another to a channel without needing to re-bracket the sequence.

\begin{figure}
\includegraphics[scale=0.75]{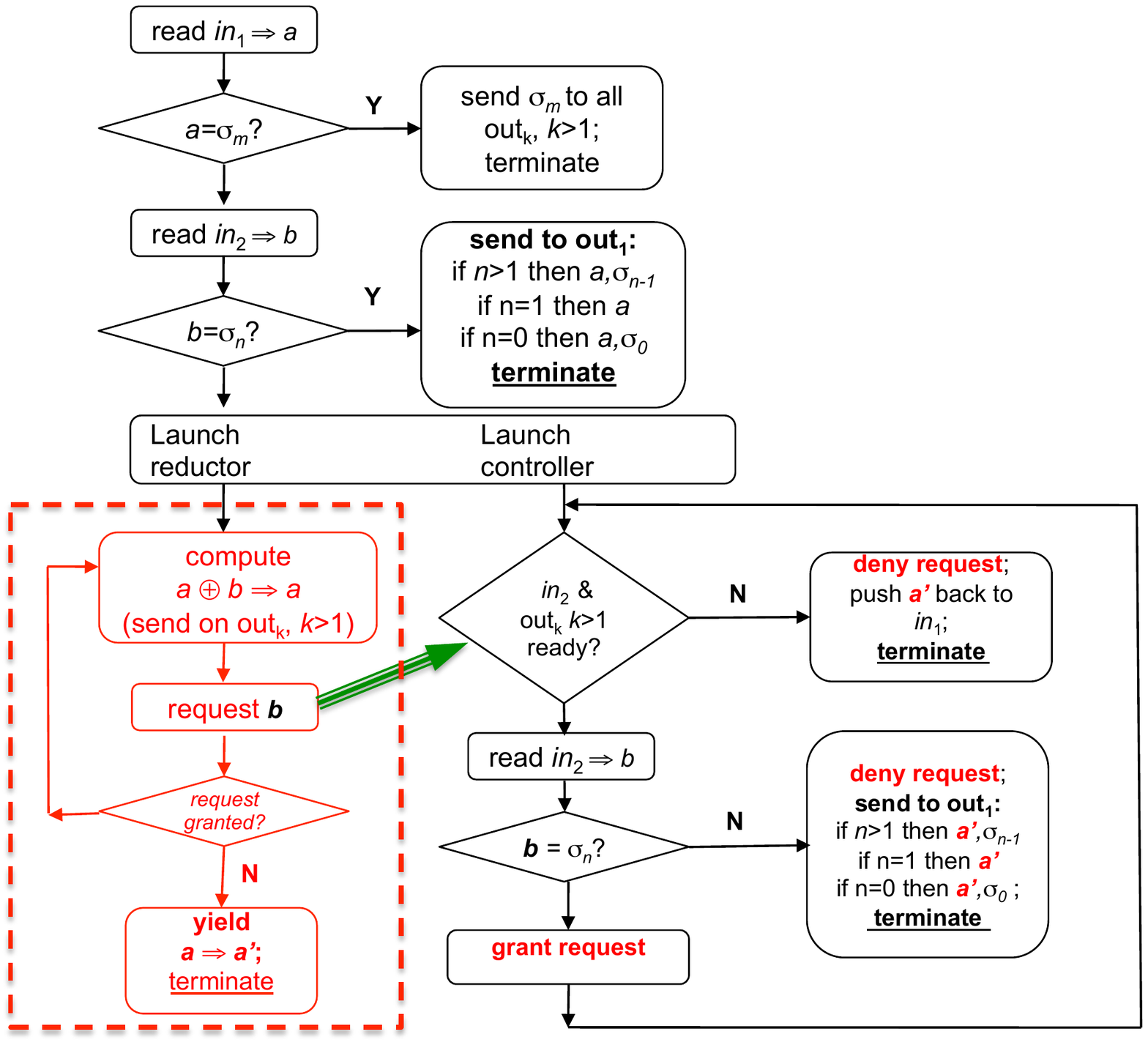}\\
\vspace{-1in}
\caption{\ak\ protocol for a reductor box. The protocol is run whenever $in_1$ and $in_2$ are ready and
all the $out_k$ are unblocked. $out_1$ must have space for a segmentation mark in addition to a data message.
\label{fig:reductors}}
\end{figure}

Let us now focus on the process of reduction (see figure  \ref{fig:reductors}). The box implements its $\oplus$ operation, and is run, as was already mentioned, when both an $a$ message and a $b$ message are available from the inputs, provided that the output channels of the reductor are all unblocked.
The box computes the result of one $\oplus$, which is a message of the same nature as $a$ to be combined with the next $b$.
The box requests the next $b$ via the Box-API, and if the request is granted, carries on and produces the new result, etc.
At some point the request will be denied
due to either (i) a segmentation mark having been received from the second stream, (ii) no further $b$ is available yet, or (iii) one of the output channels (channel 2 or above) is blocked\footnote{remember that the readiness of output channel 1 is a precondition to the activation of a reductor; once the channel is ready it cannot become blocked before a message is output to it}. In any of the three cases the box must yield the current result and terminate. In case (i) the coordinator sends the result message to the first output channel of the box. In cases (ii) and (iii) the message on the
first input channel is replaced by the result message. Consequently in case (ii) the box will be relaunched as soon as a $b$ becomes available,
using the latest intermediate result as $a$. This means that the termination of the box in cases (ii) and (iii) is completely transparent as far as the output values are concerned and is merely a form of progress control.

As the attentive reader will have noticed, a reductor can have more than one output channel. The first output channel is reserved for the results of the reduction operation and it behaves as described above.
The reductor's behaviour on the other output channels, if any, is the same as that of an inductor with respect
to the first input channel. namely, if at any time during the reduction any one of the output channels becomes blocked, then the reduction is stopped upon completion of the current intermediate step and the intermediate result replaces the $a$ message, same as in the case of inductor. When the pressure on the output drops to subcritical, the box is relaunched in the same manner as above and will proceed as if the blockage had never happened. Consequently, it is part of the reductor's contract with the TPL that any and all output messages of the reductor can only depend on the latest intermediate result (or the original value of $a$ if no intermediate result has yet been computed) and the latest received value of $b$. In this sense, a reductor is as stateless as an inductor. Both have intermediate state which can be externalised if the coordinator so demands, and both forbid any output to depend on anything but the latest intermediate state.
The bracketing behaviour for the output channels other than the first follows from the induction nature of those channels:
in the same way as it happens in the case of inductor the segmentation marks from the first input channel (which were
ignored for the reduction output), are transferred to each of the rest of the output channels: any $\sigma_k$ with $k>0$
is sent across as $\sigma_{k+1}$.

It should be noted that the reductor protocol is run whenever its preconditions are satisfied: availability of messages on both input channels and the unblocked status of the output channels. The precondition becomes potentially satisfiable (and hence checkable) whenever a new message is received on an input channel or when a message is consumed off a blocked output channel.

\paragraph{Example.}

Consider the operation $a\oplus b$ to be addition of a two-bit unsigned integer $a$ with a one-bit unsigned integer $b$. the result is the sum, $s=a+b \,{\rm mod\,} 4$ and a one-bit unsigned integer carry $c = \lfloor (a+b)/4\rfloor$. Now assume the reductor {\bf sum} uses $\oplus$ and has, accordingly, four channels: $a$, $b$, $s$,$c$, which are the first and second input and the first and second output channels, respectively. Let us set the following inputs to the two input channels:
\begin{eqnarray}
a & = & (2,1),(2) = 2, 1, \sigma_1, 2 \sigma_0 \nonumber\\
b & = & (((0,1,0))((1,1,1)(0,0,1))) = 0,1,0, \sigma_2, 1,1,1,\sigma_1, 0,0,1,\sigma_0 \nonumber
\end{eqnarray}
Upon termination of {\bf sum} the output channels will have the following content:
\begin{eqnarray}
s & = &  ((3)(0,3)) = 3,\sigma_1, 0,3,\sigma_0    \nonumber\\
c & = & ((0,0,0),(0,0,1)),((0,0,0)) = 0,0,0,\sigma_1,0,0,1,\sigma_2,0,0,0,\sigma_0 \nonumber
\end{eqnarray}

\paragraph{Monadic reductors.} A monadic reductor has one input channel from which it receives the sequence to be reduced. The reduction is based on an operator
$\oplus:\tau\to\tau\to\tau$, which is a particular case of the dyadic reductor with $\tau_a=\tau_b=\tau$.  The behaviour of the monadic reductor is similar as well.
The main difference is that the reductor is run when two messages are received. If the first message is a segmentation mark this indicates reduction of the empty sequence which leads to no output, and the protocol terminates. Otherwise the first message is a data message. If the second message is a segmentation mark, the first message is the reduction result: all monadic reductors when applied to a singleton sequence are automatically assumed idempotent. Consequently the message is passed on to the first output channel and the rest of the output channels receive no output; the protocol terminates. Finally, if two data messages are received, the box is launched, provided that none of the output channels is blocked. The box receives two messages through the API and applies the $\oplus$ operator to them, works out the result, and sends any messages it needs to send on the output channels other than the first one (i.e. on the induction channels). Then it attempts to read one further input message and to combine it with the result just obtained, etc. Similar to the case of the dyadic reductor,  the monadic one can be informed by the API that there is no further input message to reduce. That could mean that the sequence has indeed been processed, or that one of the inductor channels is blocked. Either way, the reductor must yield its current result message and terminate. It is part of the monadic reductor contract that when relaunched with the intermediate result prefixing the rest of the input sequence, the box must operate as if the termination had not happened.

\subsection{Box categories and channel names\label{sec:category}}

At the TPL data properties (such as value types) are not captured and the only features of box that are recognised are the box category and the number of the output channels. There exist seven box categories:
\begin{description}
\item[Transductor], denoted as $n$\verb$T$, e.g. 3\verb$T$ is a transductor box with three output channels.
\item[Inductor],  denoted as $n$\verb$I$, e.g. 2\verb$I$ is an inductor with two output channels.
\item[Reductor: Dyadic Ordered], denoted as $n$\verb$DO$, e.g. 1\verb$DO$ is a dyadic reductor; it has two input channels and output channel. The reduction operator is applied to the messages in the order in which they arrive on the second input channel.
\item[Reductor: Dyadic Unordered] , denoted as $n$\verb$DU$. Same as $n$\verb$DO$, except the reduction operator can be applied to the messages on the second channel in any order without affecting the result.
\item[Reductor: Monadic Ordered], denoted as $n$\verb$MO$. Same as $n$\verb$DO$, except the reduction is monadic, and consequently there is only one input channel.
\item[Reductor: Monadic Segmented], denoted as $n$\verb$MS$. Same as $n$\verb$MO$, except the list of messages on the input channel can be segmented into arbitrary sublists (without changing the order of the elements). Then the reduction operation can be applied to each sublist replacing it with the reduction result. The new list will be treated exactly like the old one, i.e. can be segmented and partially reduced, etc. until the final result is computed. An \verb$MS$ reductor must ensure that the result does not depend on the segmentation. This is usually the case when the reduction operation
is associative (but it does not need to be commutative).
\item[Reductor: Monadic Unordered] , denoted as $n$\verb$MU$. Same as $n$\verb$MO$, except the reduction operator can be applied to the messages in any order without affecting the result.
\end{description}

Boxes do not name their channels. The TPL on the other hand, needs channel names in order to be able to wire up a network, see section \ref{sec:wiring} below. When a box name is used in a wire expression, the names assigned by default are \verb$_1$ (and \verb$_2$ for a dyadic reductor) for the inputs, and \verb$_1$, \verb$_2$, etc., for the outputs. An \ak\ programmer can change that for a given box by using the renaming brackets\footnote{which is a syntactic construct not to be confused with segmentation marks: the former is a modifier of an \ak\ vertex and the latter is a special type of message}:
\begin{lstlisting}
<a,b | my_dyadic | c >
\end{lstlisting}
Here the box \verb$my_dyadic$ has two input channels and one output channel. The input channels that would have been named \verb$_1$ and \verb$_2$  without the renaming brackets are now named \verb$a$ and \verb$b$ and the output channel \verb$_1$ is named \verb$c$.

Boxes are subtyped at the TPL on the number of the output channels. A box that is expected to provide $n$ output channels can in fact have $k<n$, in which case channels numbered $k+1, k+2,\ldots, n$
are added to the box, and any time they are under negative pressure, a $\sigma_0$ is sent on them by the coordinator. Consequently the list of channel names in the right-hand side of the remaning brackets can be arbitrarily long, but if it is too short for the number of output channels present in the box, the original names not covered by the list, such as \verb$_2$, remain unchanged. A similar tactic is applied to the left-hand renaming bracket. There only one name is required usually, and it can be omitted, but in the case of a dyadic reductor, the second name may be present, as in the above example, or if not, then \verb$_2$ is retained.

The name lists on either side of the brackets can be composed of name-value pairs. For instance,
the above example can be written as
\begin{lstlisting}
<_1=a,_2=b | my_dyadic | _3=c >
\end{lstlisting}
but it is not possible to mix both forms within the same list. Finally the renaming brackets(in positional, not key-value form) can be used on their own, without the box they envelop, in which case the box name is replaced by the tilde \verb$~$. This represents a nondeterministic merge of the input channels into a single stream, which is copied to all the output channels, e.g.
\begin{lstlisting}
<a,b,c|~|d,e >
\end{lstlisting}
merges channels  \verb$a$, \verb$b$ and \verb$c$ into a stream, which is delivered to \verb$d$ and \verb$e$.

\section{Synchronisers\label{sec:synch}}

A synchroniser is a vertex that produces output messages based on a potentially unlimited history of input messages read from its input channels. In order to do this, a synchroniser maintains an internal state and makes state transitions. The states and transitions between them define which channels are read and in what order depending on the channel status (available, not available) and optionally the content of the messages.
Messages received in various states can be stored in the synchronisation storage with the single purpose to retrieve them in another state and to use them in output messages. From the mathematical point of view a synchroniser is a pair $(\Phi,\Pi)$.

$\Phi$ is a nondeterministic finite state machine $\Phi=(A,S,T)$, where $S\supseteq \{s\}$ is a set of abstract states, $s$ the start state, $A\subseteq C\times P$ the alphabet of events and $T:A\times S\to S$ is a transition matrix. Here the set $C$ is the set of the synchroniser's input channels, and the set $P$ consists of predicates on channel messages. An event $(c,p)\in A$ represents the reception of a message on channel $c$ that satisfies predicate $p$ (which can be a tautology).

\newcommand\pfun{\mathrel{\ooalign{\hfil$\mapstochar\mkern5mu$\hfil\cr$\to$\cr}}}

$\Pi: S\times\Omega\pfun V $, where $\Omega$ is the set of output channels and $V$ is the set of message values, is a path functional that defines the synchroniser output, which could be nil on any given channel in any given state. In a state $s_k$ the functional is based on the retrospective sequence of transitions from the most recent visit to the start state to $s_k$:
\[
(s_0,c_0),(s_1,c_1)\ldots (s_k,c_k)\,,
\]

where $s_0=s$ and each $c_i\in C$ is the channel that caused the transition from the state $s_i$. Denote the message received in that transition
as $\mu_i$. Then $\Pi (s_k,\omega_m) = \psi_{\sqcap} \{\mu_i \mid \rho_{ki}^m (s_i), 0\le i\le k\}$, where $\rho_{ki}^m$ is a selection predicate that defines $\Pi_k$ and the operator $\psi_{\sqcap}$ coerces the messages in the operand set to their joint greatest subtype.  The latter requires further explanation, which is given next.

\subsection{Synchroniser code\label{sec:sync}}

\begin{figure}
\begin{framed}
\small
\begin{grammar}
[(colon){$\rightarrow$}]
[(semicolon)$|$]
[(comma){}]
[(period){\\}]
[(quote){\begin{bf}}{\end{bf}}]
[(nonterminal){$\langle$}{$\rangle$}]
<synchroniser>:"synch",<name>,[<confs>],<params>,"\{"<decls>,<trans>"\}".

<confs>:"{\tt [}",<conf>,[",",<conf>]*,"{\tt ]}".
<conf>:<id>

<params>: "(",[<inparam>,[",",<inparam>]*],"{\tt |}",[<outparam>,[",",<outparam>]*],")".

<inparam>:<chan>,[":",<indepth>].
<chan>:<id>.
<indepth>:<var>;<const>.
<var>:<id>.

<outparam>:<chan>,[":",<depth-exp>].
<depth-exp>:<const>;<var>;<var>,"{\tt+}",<shift>;<var>,"{\tt -}",<shift>.
<shift>:<const>;<conf>.

<decls>:[<store-decl>;<state-decl>]*.
<store-decl>:"store",<var>,"{\tt :}",<chan-tail>,[<var>,"{\tt :}",<chan-tail>]*.
<state-decl>:"state",<type>,<var>[",",<var>]*.
<type>:"int(",<const>,")";"enum(",<id>,[",",<id>]*,")".
<chan-tail>: <chan>;<tail>.

<trans>:<tran>[",",<tran>]*.
<tran>:[<label>,":"],[<on-clause>],[<do-clause>],[<send-clause>],[<goto-clause>].

<on-clause>:"on",<chan-cond>;"elseon",<chan-cond>.
<chan-cond>:<primary>["{\tt \&}"<guard-exp>].
<primary>:<chan>,[".",<secondary>].
<secondary>:"{\tt@}",<id>;"{\tt else}";"?",<id>;["?",<id>]<pattern>.
<pattern>:"(",<id>,[",",<id>]*,["{\tt||}",<tail>, ")".
<guard-exp>:<int-exp>.
<tail>:<id>

<do-clause>: "do",<assgn>,[",",<assgn>]*.
<assgn>:<id>,"{\tt :=}",<int-exp>; <id>,"{\tt :=}",<data-exp>.

<send-clause>: "send", <dispatch>,[",",<dispatch>]*.
<dispatch>: <msg-exp>,"{\tt =>}", <chan>.
<msg-exp>: "{\tt @}",<int-exp>; "?",<id>; ["?",<id>],<data-exp>;"nil".
<data-exp>: <primary-mes>; "(",<primary-mes>, ["," <primary-mes>]*,")".
<primary-mes>: <var>;<var>=<int-exp>;"this".
<goto-clause>:"goto",<label>.

\end{grammar}
\end{framed}
\caption{The syntax of the \ak\ synchroniser\label{synch-syntax}}
\end{figure}

\paragraph{Memory.}As stated above, the synchroniser is fully defined by two functions: the transition matrix $T$ and the selection predicate $\rho$. The synchroniser thus encodes both objects in a certain structured way. Starting with $T$ we remark that the state machine can have a regular structure whereby many
transitions can be defined at once by a formula with some limited range integer variables. For example,
a machine with 8 states could have a transition matrix defined thus: $s_k \to s_{k+1\,\mbox{mod}\,8}$. In order to be able to use regular structures, \ak\ allows synchronisers to declare state variables.

The representation of $\Pi$ is also straightforward. In a given state $k$ for each output channel $\omega_m$
we note all $i$ on which $\rho_{ki}^m$ is true. Those message values must be stored in a previous state
and recalled in state $k$. It is expected that the Boolean vector $w_i=\rho_{ki}^m$ has only very few
true elements. Consequently the storage mechanism that \ak\ provides for synchronisers is in the form
of individual {\em store variables}, which are associated with input channel types. Thus a declaration
such as $x:\mbox{foo}$ declares $x$ to be of the same type as the messages coming from the input
channel foo. When a message is received on a channel, it can be matched with a pattern in order to extract integer parameters needed to select a specific transition. Such a message can be stored either as is (referred to by the keyword {\tt this}) or without the parameters, in which case the pattern must declare a name for the message tail. Tail names can be used in declaration of store variables in the same manner as channel names.
A store variable can be assigned messages from several channels: $x:\mbox{foo,bar}$,
in which case it is assumed that the variable has the least common type that the channels can be
coerced up to. More about typing considerations can be found in the next chapter. Types are the concern of the Constraint Aggregation Layer, CAL, where full support is given to logic mechanisms
that determine every piece of the typing information required to compile the coordination program and specialise the boxes. At the TPL all we need to know about type is specific issues of data extraction from
messages and of combining one or more messages with named, integer-valued entities into a single output message.

\paragraph{Syntax.} The formal syntax of the \ak\ synchroniser is given in figure \ref{synch-syntax}. In this section we provide informal descriptions and examples
to facilitate the understanding of its basic structures and mechanisms. We start with a binary zip example,
see figure \ref{zip2}.

\begin{figure}
\lstset{numbers=left, numberstyle=\small, stepnumber=1, numbersep=8pt}
\begin{lstlisting}[frame=single]
synch zip2(a:0,b:0 | c:0)
{
	store ma:a, mb:b;
	
start:	on a do
		ma:= this
		goto s1;
	on b do
		mb:= this
		goto s2;

s1: 	on b send (ma,this) => c 	goto start;
s2:	on a send (mb,this) => c 	goto start;
}
\end{lstlisting}
\caption{Synchroniser {\tt zip2}\label{zip2}}
\end{figure}
Here line 1 declares zip2 as a synchroniser with input channels \verb$a$ and \verb$b$ and the output channel \verb$c$.
The channels \verb$a$ and \verb$b$ are required to have the bracketing depth 0, while the channel \verb$c$ is guaranteed
to have the bracketing depth 0.
Line 3 declares two store variables, \verb$ma$ capable of storing a data message from channel \verb$a$ and \verb$mb$
capable of storing a data message form channel \verb$b$.

Lines 5--10 define the behaviour in the start state . Remember that each synchroniser has a start state and that
whenever the state is reached again, all store variables lose their associated values. The \verb$on$ clause
defines the condition on which the transition takes place. The channel name on its own stands for the availability
predicate for the corresponding channel, i.e. the condition that a message of any kind is available. The \verb$do$
clause is an action list that contains actions that evaluate the functional. The \verb$send$ clause forms and sends
output messages.
Finally the \verb$goto$ clause
defines the state transition. It is easy to see that the synchroniser in its start state accepts messages from channels $a$
and $b$, stores them in store variables \verb$ma$ and \verb$mb$, respectively,  depending which channel is available and then
transitions to the states \verb$s1$ and \verb$s2$ respectively.

Line 12 defines the behaviour in the state \verb$s1$. The synchroniser finds itself in this state when it has received a message
on channel \verb$a$ and stored it in \verb$ma$. In this state the synchroniser can only receive messages from channel \verb$b$ since
there is no transition triggered by channel \verb$a$ and defined in this state. Hence channel \verb$a$ at this point is blocked
by the synchroniser. When the message on channel \verb$a$ is received, the send clause computes the concatenation of the
current message \verb$this$ and the content of \verb$ma$, and sends the result to channel \verb$c$.

Line 13 defines similar behaviour in state $s2$.

Since none of the transitions tests for the end-of-stream condition, i.e. $\sigma_0$, the default rule applies: the receipt of
a $\sigma_0$ causes the synchroniser to terminate after sending $\sigma_0$ to all its output channels.   Also notice
that in the state \verb$start$ both input channels may be ready, but a state machine receives input symbols one at a time.
Which transition will be triggered under such circumstances is defined by the fairness policy: the coordinator will
ensure that when more than one transition is possible in a given state, all choices will be made with the same frequency.
\ak\ supports an alternative fairness policy as well: prioritised alternation whereby the \verb$on$ clauses
are prioritised top-down. This is indicated by using the \verb$elseon$ keyword instead of \verb$on$ wherever the subsequent
choice is only taken when the preceding one(s) have failed.

The next example is a running counter, see figure \ref{counter}.
\begin{figure}
\begin{lstlisting}[frame=single]
synch counter (a, c | b, error)
{
	state int(8) count;

start:		do count:=0 goto work;

work: 		on a & count < 255 do
		count := count+1, cnt =count
		send (cnt, this) => b
		goto work;
		
		on a & count=255 do
		count := 0, errcode: = -1, cnt:=0
		send errcode => error,   (cnt, this) =>b
		goto start;
		
		on c.(cnt) do count:=cnt;
}
\end{lstlisting}
\caption{Synchroniser {\tt counter}\label{counter}}
\end{figure}
Here we have one input and two output channels, all three unsegmented. We use one state variable \verb$count$ defined as an 8-bit
integer on line 3, which is initialised with 0 in the initial state (line 5). The counter is incremented every time a message is
received, except when it reaches 255, in which case an error message is sent
down the channel {\em error}.
Finally, the channel $c$  must carry messages that contain the counter \verb$cnt$. When received, the corresponding value is assigned to the state variable \verb$count$ and the synchroniser transitions back to (i.e. remains in) its current state \verb$work$. Notice that \verb"errorcode" has not been defined as either state or store variable. All such variables are considered aliases for integer expressions; they need not be declared.

Let us now see what happens when channels have nonzero bracketing depths. The following example (figure \ref{listMerge}) is a priority merger,
which takes two bracketed streams of the same depth \verb$d$: \verb$a$ and \verb$b$, and joins them into a stream \verb$c$ of depth
\verb$d+1$ at the top level of brackets. The result is a stream of pairs, the first member of each pair taken from the
top level (depth \verb$d$) sublist of \verb$a$ and the second member from \verb$b$.
\begin{figure}
\begin{lstlisting}[frame=single]
synch listMerge (a:d, b:d | c:d+1)
{

start:		on a.@k & k=d send this => c goto alt;
		on a.else send this => c goto start;

alt:		on b.@k & k=d send @k+1 => c goto start;
		on b.else send this => c goto alt;
}
\end{lstlisting}
\caption{Synchroniser {\tt listMerge}\label{listMerge}}
\end{figure}
Notice on line 1 both channels are declared to have depth $d$ which creates a constraint\footnote{channel depth constraints are handled by TPL itself, since they are essential for channel and synchroniser behaviour} that is resolved between the synchroniser
and the vertices responsible for the production of the two streams, giving $d$ some value. On line 4 the condition \verb$a.@k$
checks that the channel \verb$a$ is ready and holding a segmentation mark $\sigma_k$
(\verb$k$ is instantiated if it does, otherwise the condition fails) and also the value of $k$ is tested. The condition $a.else$ is equivalent to $a$
except it is tested {\em after} any other condition involving the channel. It is important to understand that although several different
channels can be tested in any given state, once a test has established the readiness of a channel, the synchroniser is committed, hence
the set of conditions applied to the message on any input channel $x$ must be exhaustive. If it is not, the final clause
\begin{lstlisting}[numbers=none]
		on x.else;
\end{lstlisting}
will be assumed, which discards the input message and transitions the synchroniser back to its current state.

On line 7, the segmentation mark $\sigma_{k+1}$ is sent to channel $c$.

\paragraph{Configuration parameters.} A synchroniser may make use of various integer constants in defining transition between states. It is convenient to be able to change those constants without having to provide a trivially altered synchroniser program. Configuration parameters occur elsewhere in \ak\ and serve the same purpose; those occurrences will be discussed later.

The ``running counter'' example modified for the arbitrary size counter is shown in figure \ref{cnt1}

\begin{figure}
\begin{lstlisting}[frame=single]
synch counter [bits] (0, 0 | b, error)
{
		state int(bits) count;

start:		do count:=0 goto work;

work: 		on a & count < 2^bits-1 do
		count := count+1, cnt=count
		send (cnt,this) => b
		goto work;
		
		on a & count=2^bits-1 do
		count := 0, errcode=-1, cnt=0
		send errcode  => error,   (cnt,this) =>b
		
		on c.(cnt) do count:=cnt;
}
\end{lstlisting}
\caption{Configurable version of the synchroniser in figure \ref{counter}\label{cnt1}}
\end{figure}
The configuration parameter \verb$bits$ is specified in square brackets before the channel signature.
A coma-separated list of parameters can be used when more than one is required. Finally, a parameter can control the depth of a channel, and can even make a channel disappear from the list (which is indicated by a depth of -1). Such ``cancelled'' channels must not have data sent to them in the synchroniser program; the programmer is responsible for introducing an appropriate configuration parameter in the conditional part of the relevant transitions. Observe that the input channels to a synchroniser never present such a problem, since input channels disappear when they receive a $\sigma_0$.

\paragraph{Integer expressions and message structure.} Finally, let us comment on the nature of the nonterminal $\langle\hbox{int-exp}\rangle$ occurring in the syntax in figure \ref{synch-syntax}. This stands for the C-style arithmetic expression with all the standard unary and binary operators.

A message at an input of a synchroniser is interpreted as a set of named entities associated with either integer or unknown values. The synchroniser can name some or all of those values, provided that they are associated with integers, and can also name the message comprising the rest of the entities (see ``tail'' in figure \ref{synch-syntax}). The named entities can be mentioned in any integer expressions, and the tail can be stored in a store variable for the corresponding channel. If the synchroniser does not require access to the content of the message it receives in a particular state, it refers to it by the keyword {\bf this}.

At the output messages can be composed from various sources: the message just received, i.e. {\bf this} is one such, but also stored messages and named integer values. Those are assembled into a single parenthesised list. To support message formats where several variants of a message are possible, which should be distinguished for the purposes of synchronisation, a qualifier \verb"?"$\alpha$ is available as an input condition. This same qualifier qualifies output messages in send-clauses as belonging to a particular variant.

CAL checks the consistency of combined messages with the same variant name (or all of them on a given channel if variants are not used for output) and their correspondence to any input requirements at the other end of each of the synchroniser's output channels. Similarly CAL deals with input formats and their correspondence to the suppliers' output definitions.

\paragraph{Synch-table.}It is often the case that a single stream carries elements of several ``logical'' subsequences arbitrarily interleaved into a single sequence of messages. In such multi-sequential streams synchronisation often occurs inside each individual subsequence as its messages are combined   into a new subsequence of the same kind. This is somewhat analogous to applying a reductor to a stream independently to each bracketed segment of it, except in the case of a multi-sequential stream it is not segments separated by brackets but subsequences identified solely by their message markers that structure the data flow. The markers could be format-related, in which case different synchronisers can be engaged to do the synchronisation job. Alternatively, the whole multi-sequence could carry messages of the same type or types, with different subsequences being distinguished by one or more {\em indices}, i.e. integer numbers associated with certain identifiers within each message. Messages with different index values belong to different subsequences, and all messages with the same values belong to the same subsequence in the order in which they occur in the stream. In the case of indexing, all subsequences would normally require separate copies of the same synchroniser. Each copy would see only messages with one particular set of index values on all its input channels, and would produce similarly indexed output messages on all channels.

Figure \ref{synchtab-syntax} defines the syntax of a {\em synch-table}. It is an array of synchronisers, indexed by $\langle{\rm ind}\rangle$s within their declared $\langle{\rm lim}\rangle$s. Each input channel is split into subsequences corresponding to a specific  combination of the index values and delivered to the corresponding replica of the $\langle{\rm synchroniser}\rangle$. Similarly the subsequences yielded by the replicas on their output channels are interleaved into joint output channels of the $\langle{\rm synchtab}\rangle$. Note that the table does not have a separate name, being, as it were, a modifier of an ordinary synchroniser. The indices are scoped over the synchroniser included in the table.

By default, an output message contains all the indices and their values correspond to the replica that has produced it. However, a synchroniser may include an explicit name=value pair into an output message as per figure \ref{synch-syntax}. This would result in the output message being inserted in another replica's output sequence non-deterministically.

Despite the nondeterminism of the subsequence merge, not everything in the function of individual replicas is independent.
Output of a bracket $\sigma_i$ with any valid $i$ on an output channel $c$ causes barrier synchronisation of $c$ in the following way:
\begin{enumerate}
\item Further output of the replica on $c$ is buffered out or blocked, and in either case not communicated to the joint channel.
\item When all other replicas become similarly blocked, a {\em single} $\sigma_k$ is yielded by synch-table on $c$ with the value of $k$ being the minimum of all $i$ from the various $\sigma_i$ encountered in the event. Then the blocked outputs are unblocked.
\end{enumerate}
By contrast, if a bracket $\sigma_i$ is encountered on an input channel, it is broadcast to all the replicas of the synch-table.

\begin{figure}
\begin{framed}
\begin{grammar}
[(colon){$\rightarrow$}]
[(semicolon)$|$]
[(comma){}]
[(period){\\}]
[(quote){\begin{bf}}{\end{bf}}]
[(nonterminal){$\langle$}{$\rangle$}]
<synchtab>:"tab","{\tt [}",<inds>,"{\tt ]}",<synchroniser>.
<inds>:<ind>,"{\tt :}",<lim>;<ind>,"{\tt :}",<lim> ,"{\tt,}",<inds>.
<ind>:<id>.
<lim>:<id>;<const>.
\end{grammar}
\end{framed}
\caption{The syntax of the \ak\ synch-table\label{synchtab-syntax}}
\end{figure}

\section{Wiring\label{sec:wiring}}

The act of connecting (atomic or compound) vertices into a network is called {\em wiring}. In \ak\ wiring is performed by an algebraic expression
with networks as operands and wiring patterns as operators. The operands can be either primitive networks, i.e. boxes
and synchronisers, or wiring expressions in their own right. Hence \ak\ wiring is hierarchical. We will now focus on
the wiring patterns, looking into one at a time and in the next section will define the \ak\ program structure.

\subsection{Formal definitions}

A vertex of a streaming network can be abstracted as a triple $v=(L_v, I_v, O_v)$, where $L_v$ is a label (that denotes
the function of the vertex $v$), $I_v$ is the set of input channels to $v$, and $O_v$ is the set of its output channels.
Two vertices are wired together when an input channel of one is identified withe an output channel of the other.
A network $N:\mathcal{N}$ can now be defined as a quadruplet $N=(\mathcal{V}_N, \hbox{\bf w}_N, \mathcal{I}_N, \mathcal{O}_N)$,
where $\mathcal{V}$ is a set of vertices, $\hbox{\bf w}_N\subseteq \mathfrak{I}_N\times\mathfrak{O}_N$ is a wiring relation
between the sets of input an output channels
\begin{eqnarray*}
\mathfrak{I}_N &=& \bigcup_{v\in\mathcal{V_N}} I_v \nonumber\\
\mathfrak{O}_N &=& \bigcup_{v\in\mathcal{V_N}} O_v \nonumber\\
\end{eqnarray*}
and
$\mathcal{I}$ and $ \mathcal{O}$ are the set of input and output channels of the net as a whole, respectively.
In particular a vertex can be made into a singleton network: $sing(v) = (\{v\},\emptyset, I_v, O_v)$

Note that a wiring may wire a single output channel to more than one input channel
and a single input channel to more than one output channels. The semantics of the former is {\em copying}: each message output on the channel will be received by each of the input channels that the output channel is wired to. The semantics of the latter is {\em merging}: when more than one output channel is wired to a single input channel, the messages from the output channels are transferred to a single input channel in no particular order, i.e. nondeterministically. It is also possible for a channel to merge several inputs and copy the stream to several outputs. Consequently the wiring relation is completely generic: it can leave channels unwired and can also lead to one-to-one, one-to-many/many-to-one or many-to-many connections.  The channels in \ak\ are named; accordingly we introduce the naming function $\iota: \mathcal{C}\to \mathcal{D}$, where $\mathcal{C}$ is the set of all channels in the program and $\mathcal{D}$ is a set of valid identifiers.
For any $c\in \mathcal{C}$, $\iota(c)$ denotes the identifier of the channel.

A valid network $(\mathcal{V}, \hbox{\bf w}, \mathcal{I}, \mathcal{O})$ must satisfy the following {\em consistency conditions}:
\begin{description}
\item[Concealment]  \[(\forall  (c,c^\prime)\in {\bf w}) c\notin \mathcal{I}\wedge c^\prime\notin\mathcal{O}\]
\item[Completeness] \[(\forall v\in \mathcal{V}, c\in I_v) \exists c^\prime: (c,c^\prime)\in {\bf w} \vee c\in \mathcal{I}\]
\[(\forall v\in \mathcal{V}, c\in O_v) \exists c^\prime: (c^\prime,c)\in {\bf w} \vee c\in \mathcal{O}\]
\[(\forall c\in\mathcal{I}\,\exists v\in\mathcal{V}) c\in I_v\]
\[(\forall c\in\mathcal{O}\,\exists v\in\mathcal{V}) c\in O_v\]
\item[Identity]
\[(\forall  (c,c^\prime)\in {\bf w})\iota(c)=\iota(c^\prime)\]
\end{description}

Informally, the concealment condition forbids the channels that connect vertices to be members
of the input and output sets; the completeness condition demands that each channel that is not connected to another
vertex contribute to the network input/output channel sets and vice versa; and finally the identity condition
stipulates that the wiring only connects identically named channels.

A wiring function $P_1: \mathcal{N}\to\mathcal{N} $ that applies additional wiring to an existing network can
always be defined in the form: \[
P_1 (\mathcal{V}, \hbox{\bf w}, \mathcal{I}, \mathcal{O}) = \left(\mathcal{V}, \hbox{\bf w}\cup \hbox{\bf p},
\mathcal{I}\setminus\hbox{dom}(\hbox{\bf p}), \mathcal{O}\setminus\hbox{img}(\hbox{\bf p})\right)\,,
\]
which automatically satisfies the consistency conditions for any relation $\hbox{\bf p}\subseteq \mathcal{I}\times \mathcal{O}$
provided that for any $c\in \mathcal{I}, c^\prime \in \mathcal{O}$, $c\,\hbox{\bf p}\, c^\prime \Rightarrow \iota(c)=\iota(c^\prime)$.
In the sequel, the relation $\hbox{\bf p}$, and, where no ambiguity would arise, the function $P_1$, will be called a {\em wiring
pattern}.  A wiring function can also be written for the case of two operand networks,
$P_2: \mathcal{N}\times\mathcal{N}\to\mathcal{N}$, thus:
\[
P_2 \left( (\mathcal{V}^{[1]}, \hbox{\bf w}^{[1]}, \mathcal{I}^{[1]}, \mathcal{O}^{[1]}),
(\mathcal{V}^{[2]}, \hbox{\bf w}^{[2]}, \mathcal{I}^{[2]}, \mathcal{O}^{[2]})\right) =
(\mathcal{V}, \hbox{\bf w}, \mathcal{I}, \mathcal{O})\,,
\]
where
\begin{eqnarray*}
\mathcal{V} &=& \mathcal{V}^{[1]}\cup\mathcal{V}^{[2]}\\
\hbox{\bf w} &=& \hbox{\bf w}^{[1]} \cup \hbox{\bf w}^{[2]}
										\cup \hbox{\bf p} \nonumber\\
\mathcal{I} &=& \mathcal{I}^{[1]}\cup\mathcal{I}^{[2]}\setminus\hbox{dom}(\hbox{\bf p})\nonumber\\
\mathcal{O} &=& \mathcal{O}^{[1]}\cup\mathcal{O}^{[2]}\setminus\hbox{img}(\hbox{\bf p})\nonumber\\
\end{eqnarray*}
using a wiring pattern $\hbox{\bf p}\subseteq \mathcal{I}\times \mathcal{O}$ that satisfies the above identity condition
\[(\forall c\in \mathcal{I}, c^\prime \in \mathcal{O}) c\,\hbox{\bf p}\, c^\prime \Rightarrow \iota(c)=\iota(c^\prime)\,.\]

\subsection{Serial, parallel and wrap-around connections\label{sec:series}}

\ak\ provides a small set of wiring patterns which is sufficient to achieve arbitrary wiring of the vertices. To start with,
let us confine ourselves to the case of an acyclic graph.

The \textbf{\emph{ serial connection}} is denoted $N_1\hbox{\bf ..}N_2$ and is defined by a wiring function $P_2$ (from the previous section)
with the following wiring pattern ${\bf p}$:
\[
c\,\hbox{\bf p}\,c^\prime \equiv \iota(c)=\iota(c^\prime) \wedge c\in\mathcal{I}^{[2]}\wedge c^\prime\in \mathcal{O}^{[1]}\,.
\]
Informally, all outputs of the first operand are wired to identically named inputs of the second operand if they exist. The rest of the channels
contribute to the input/output sets of the result network, by construction of $P_2$.

The \textbf{\emph{ parallel connection}} is denoted $N_1||N_2$ and is defined by the wiring function $P_2$ with the empty wiring pattern ${\bf p}=\emptyset$. Informally, the two operand networks are placed side by side without connection and their input and output channels form
the input and output channel sets of the result.

The \textbf{\emph{ wrap-around connection}} has more than one version. The simpler version is denoted $N\backslash$ and is defined by the wiring function $P_1$ from the previous section, with the following wiring pattern
\[
c\,\hbox{\bf p}\,c^\prime \equiv \iota(c)=\iota(c^\prime) \wedge c\in\mathcal{I}\wedge c^\prime\in \mathcal{O}\,.
\]
Here each output channel of the operand that matches an input channel by name is wired to it, thus completing a cyclic connection.
Such channels differ from channels that are not used in a wrap-around connection by their behaviour with respect to progress. A wrap-around channel is {\em depressurised}, i.e. it is unable to block its input end if messages are not read off the output end. Consequently the wrap-around channel will queue up its input messages without limit; this may eventually cause a terminal error if the amount of memory available for the queue proves insufficient. The reason for depressurisation of the wrap-around channels is the fact that  a cyclic connection, if pressurised, may cause a deadlock.

In a cyclic network it is often the case that wrap around links are fed with data functionally dependent on  specific input (not wrap-around) channels. In such circumstances the growth of the queue in a wrap around channel could be limited by transferring the back pressure over to
such input channels. This is written as $N\backslash(-c_1,..,c_k)$, where $c_1,..,c_k$ are a subset of the $\mathcal{I}_N$ to which the back pressure from each wrap around channels is transferred.

Finally, the network $N$ may need to have identically named (but not wired together) input and output channels. To prevent them from being identified automatically as a single wrap-around channel, they can be specified explicitly in the construct as follows:  $N\backslash(r_1,..,r_m)$, where $r_1,..,r_m$ are names of the channels to be identified in the wrap-around fashion, while the rest of the channels will contribute to the input/output sets of the result instead. The wiring pattern is modified accordingly:
\[
c\,\hbox{\bf p}\,c^\prime \equiv \iota(c)=\iota(c^\prime) \wedge c\in\mathcal{I}\wedge c^\prime\in \mathcal{O}\wedge \iota(c)\in\mathcal{R}\,,
\]
where $\mathcal{R}$ is the set of channel names $r_1,..,r_m$ specified in the connection.

Both pressure transfer channels and the wrap-around channels can be specified at the same time, in which case they are placed on the opposite sides of the minus sign. The wrap-around channel name list can optionally be preceded by a hat \verb$^$ in which case the list is treated as an exclusion list, i.e. all channels {\em but} the ones specified are considered for wrapping around. Figuring out the wiring pattern in this case is left as an exercise to the reader.

\subsection{Arbitrary topology}

The three connections above are sufficient to achieve an arbitrary graph topology of the channels. Consequently an arbitrary KPN can be wired in \ak\ by using suitable vertices, choosing suitable channel names to ensure the identity of input/output channels and applying a combination of connections of those three kinds.

To see that this is the case, let us consider an arbitrary labelled, directed graph $G=(V,{\bf r}, L)$, where $V$ is a set of vertices, ${\bf r}\subseteq V\times V$ a set of edges and $L: V\times V\to D$ the edge labelling function. The input and output channels of a network are modelled in the graph by two marked vertices $I$ and $O$, to which the input and output edges, respectively, are incident. The vertex $I$  has no incoming edges and the vertex $O$ no outgoing edges. Let us also assume that all edges are named distinctly, so that no pair of edges are found having the same name and $L$ is bijective. We will now see how the graph can be transformed into an algebraic expression that uses the three fundamental connections as operators and names of vertices as operands.
\begin{description}
\item[step 1]. If the graph is cyclic, find an acyclic subgraph. \footnote{As a design consideration, since wrap-around edges are troublesome for progress control, one would do well to try to find the largest acyclic subgraph. This is known to be an NP-hard problem for arbitrary graphs, but an approximation is always available, i.e. a smaller acyclic subgraph.} The edges left out form the {\em feedback edge set}.  Break those edges and replace each such edge named $x$ by a pair of channels, one input $x$ and one output $x$ (so that their identification by name in wiring would result in the original edge). Include the input/output $x$ into the input/output set, i.e. make the input $x$ incident to  $I$, and the output $x$ to $O$. Continue breaking the edges in this manner until the graph becomes acyclic.

\item[step 2]. As the graph is now acyclic, it must have vertices with no incoming edges (such as $I$, but there may be more than one of those).
Call them graph inputs. Introduce the stage function $\gamma: V\to \mathbb{N}$, such that for any $v\in V$ $\gamma(v)$ is the length of the longest path from a graph input to $v$. Assuming that the graph is connected, a path of this kind must exist. Label each vertex with its value of $\gamma$, and form the name-sets for its input and output channels (i.e. incoming and outgoing edges incident to the vertex). Finally remove the vertices $I$ and $O$. The result is a set of labelled vertices, each having input and output channel names.

\item[step3] Introduce the following representation of the graph G as an \ak\ network:

\[
\left ((v^{[0]}_1 || v^{[0]}_2 || \ldots || v^{[0]}_{k_0})\,\hbox{\bf\Large..}\, (v^{[1]}_1|| v^{[1]}_2 || \ldots || v^{[1]}_{k_1}) \,\hbox{\bf\Large..}\, \ldots \,\hbox{\bf\Large..}\, (v^{[d]}_1 || v^{[d]}_2 || \ldots || v^{[d]}_{k_d})\right) \backslash
\]

Here all $v^{[j]}_{i}$ are the vertices that have the stage label $j$.
\end{description}

It easy to prove that there is a one-to-one correspondence between the original graph and the \ak\ network in the following sense:
wherever there exists an edge $(v,w)$ named $x=L(v,w)$ in the graph $G$, there is also a channel named $x$ wired between the corresponding vertices $v^{[j_1]}_{i_1}$ and $v^{[j_2>j_1]}_{i_2}$ and vice versa. Consequently an arbitrary KPN can be transformed into an \ak\ network
whenever the vertices of the KPN are functions expressible as boxes or synchronisers. When they are not, an \ak\ subnetwork can always be constructed, which implements an arbitrary stream-to-stream transformation. Indeed, a KPN vertex is a stream-monotonic function, which
responds with a stream of outputs to an input sequence. A synchroniser wired to a transductor with a pair of channels in each direction is able to express an arbitrary stepwise computation dependent on input streams; if all else fails, this tandem can be used to create a KPN vertex as an \ak\ subnet.

We conclude this section by remarking that as the above analysis shows, \ak\ is at least as potent as KPN, and a KPN program can be
transformed to \ak\ almost mechanically. In fact \ak\ can do things that are not possible in KPN, since the \ak\ synchroniser can also have useful nondeterministic behaviour. Another helpful feature of \ak\ is that it has provisions for homogenous wiring, which is the subject of the next section.

\subsection{Fixed-Point Series\label{sec:fps}}

In this section we will dwell on the wiring patterns based on a replicated operand network. Replication of networks
in \ak\ does not necessarily involve significant additional resources; indeed the boxes are all stateless and consequently have a zero footprint when quiescent, and the synchronisers, likewise, require no resources in their start state. Consequently the cost of replication is not felt until the replicas become active, which happens when the first message is received and only lasts until all messages have left the replica and all its synchronisers have returned to the initial state.

Strictly speaking the fixed-point series (FPS) is more than a connection since it does not simply wire the replicas of its operand. It also creates a set of output channels and in some cases augments the operand with some auxiliary vertices. However, in the main it still does wire the replicas in a serial fashion. We will consequently call FPS a ``connection''.

Consider a vertex $v$ that has an input and an output channel, both named $x$.
\begin{mydef} The vertex $v$ is said to have a {\bf forward fixed point} in $x$ if and only if the following requirements are satisfied:
\begin{enumerate}
\item There exists a condition $p(m)$ on the content of the message $m$ received by the vertex on the input channel $x$ under which it follows a unique non-branching path to the output channel $x$ without traversing any boxes.
\item The path can traverse synchronisers, but then whenever $p(m)$ is true and the synchroniser is in the start state, it must accept $m$ and transition back to the start state while sending the message $m$ on the path unchanged and without producing any other output.
\end{enumerate}
The condition $p$ may not be unique, and when it is not,  a disjunction of all such conditions is called the {\bf fixed-point condition} of the vertex on channel $x$. The condition can also be a tautology, in which case the forward fixed point is called unconditional.
\end{mydef}
\begin{mydef}
The vertex $v$ is said to have a {\bf reverse fixed point} in $x$ if and only if the following statements hold:
\begin{enumerate}
\item A unique non-branching path from the input to the output channel $x$ exists that does not traverse any boxes.
\item Every synchroniser $S_i$ on the path has a subset of states, which we denote as $s_i$, such that in each of these states every message on the path is immediately transferred without being changed or even stored, causing the synchroniser to remain in the same state\footnote{It should also be borne in mind that the values of any state variables form a part of the synchroniser state.}. In a state from $s_i$ the synchroniser $S_i$ may still be sensitive to other input channels, as long as this does not, under any circumstances, cause a transition to a state outside $s_i$.
\end{enumerate}
The vertex $v$ is said to be in a {\bf reverse fixed point state} on channel $x$ when each $S_i$ is in a state that belongs to its $s_i$.
\end{mydef}
An FPS is a form of replication wiring whereby an infinite chain of replicas is created, connected in series (see section \ref{sec:series}). The connection is denoted as $A^*$ for any operand network $A$ and can be thought of as the equivalent of
\[
A^* = A^\prime\hbox{\bf ..}A^\prime\hbox{\bf ..}A^\prime\hbox{\bf ..}\; \ldots
\]
where $A^\prime$, called the {\em streamlining} of the vertex $A$, is a network that contains $A$ and provides some additional wiring to ensure that each output channel of $A^\prime$ matches an input channel
and vice versa. We will dwell on the streamlining procedure a little further down, and at this point only remark that if all output channels of $A$ match its input channels bijectively, $A^\prime=A$.

A replica $A^\prime$ is called {\em inactive} whenever all of its synchronisers are in their start states, none of its channels has messages in them and no box is running. It is easy to see that at any given time only a finite prefix of the chain is active\footnote{on the assumption that any message-passing on a channel takes a non-zero time}. As written, the above formula does not make it clear how the output of the infinite chain can physically be produced, let alone connected to the rest of the network due to the fact that the chain is infinite. Here is how it is done.

The FPS connection $A^*$ defines the output channel-name set $\mathcal N_{out}$ as follows:
\[
{\mathcal N_{out}} = \{\iota(c)\mid  c \in {\mathcal O} \wedge \hbox{fp}(c)\}
\]
where ${\mathcal O}$ is the output channel set of $A$ and the predicate $\hbox{fp}(c)$ is true on any channel $c$ that has a forward fixed point. The FPS creates a set of fresh output channels $O^*$ taking the names from the set $\mathcal N_{out}$.   A message coming to an inactive replica on any channel $c$ with $\iota(c) \in \mathcal N_{out}$ and which satisfies the fixed-point condition on that channel is  immediately transferred to the identically named output channel from $O^*$. That is the only way output is produced from an FPS.

The reverse fixed point has a differ purpose: it optimises an input connection that has to cascade through the chain to a replica that is ready to accept the data. Any input channel $x$ wired to an active replica $A^\prime_i$ that transitions to a reverse fixed point state on that channel is disconnected from  $A^\prime_i$ and dynamically rewired to the input channel $x$ of the next replica on the chain $A^\prime_{i+1}$.

Notice that the effect of both fixed point actions is in fact transparent: the extracted messages that are transferred to the channels from $O^*$ would provably never change and would not cause any computations anywhere no matter how long they propagated along the chain. In this sense the messages "warp through" the infinite chain of replicas. Likewise the dynamically reconnected input channels could be left alone, which would only affect the performance, but not the results of the computation as it can be proven that those input messages, if not delivered straight to their destination,  would have been cascading through the chain unchanged and causing no substantive change in any computations.

The \ak\ compiler is in a position to detect both types of fixed point from the wiring of the network and the synchroniser definitions that occur in the operand.

\paragraph{Streamlining.}

The FPS connection above depends for its semantics on the fact that each replica has matching input and output interfaces. When the operand is such that its input and output interfaces do not much, the following {\em streamlining} procedure is applied to augment the operand as appropriate to eliminate the mismatch.

Let us again assume that the operand's sets of input and output channels are $\mathcal I$ and $\mathcal O$, respectively. The set of mismatched inputs and outputs are:
\begin{eqnarray*}
\bar{\mathcal I} & = & {\mathcal I}\setminus {\mathcal O}\;\;\hspace{4em}\hbox{and} \\ \nonumber
\bar{\mathcal O} & = & {\mathcal O}\setminus {\mathcal I}\,, \nonumber
\end{eqnarray*}
respectively.

Let $A^\prime =A$ initially. The streamlining procedure breaks down into two parts
that can be done in either order. First there is the streamlining of the outputs, as follows.

For each $c\in \bar{\mathcal O}$
replace $A^\prime$ by
\begin{center}
$A^\prime$\verb@..<@$\nu$\verb@|~|@$\phi$\verb@>..<@$\nu$\verb@,@$\phi$\verb@|~|@$\nu$\verb@>@.
\end{center}
where $\nu=\iota(c)$ and $\phi$ is a fresh name. The above formula binds the channel named $\nu$ to a channel named $\phi$ and then merges it with the channel named $\nu$ that is now available at both the input and the output of the new network $A^\prime$. Using the above fixed-point definitions it is easy to satisfy oneself that the new network $A^\prime$ has an unconditional forward fixed point on channel $\nu$.

Now let us streamline the inputs. For each $c\in \bar{\mathcal I}$ replace $A^\prime$ by
\begin{center}
\verb@make_rfp(@$\nu$\verb@|@$\phi$\verb@,@$\nu$\verb@)..(<@$\phi$\verb@|~|@$\nu$\verb@>..@$A^\prime$\verb@)@
\end{center}
where again $\nu=\iota(c)$ and $\phi$ is a fresh name. The vertex \verb$make_rfp$ is the following synchroniser:
\begin{lstlisting}[frame=single]
synch rfp (u:k | v:k-1, w:k)
{
start:	  on u.sigma(k)  send sigma(k-1) => v goto bypass;
	  on u.sigma(0) send sigma(0) => v, sigma(0)=>w goto start;
	  on u.sigma(m) & m < k send sigma(m-1) => v goto start;
	  on u.else send this => v goto start;

bypass: 	on u send this=>w goto bypass;
}
\end{lstlisting}
The synchroniser treats the input stream as a list of lists. The first element, which is a list in its own right is directed to channel \verb$v$. As soon as it is fully sent (i.e. upon the receipt of its closing bracket) the synchroniser switches itself to the bypass state in which the conditions of the reverse fixed point on the path consisting of the channels \verb$u$ and \verb$w$ are satisfied.

\section{Networked unit: net\label{sec:net}}

\begin{figure}[t]
\begin{framed}
\small
\begin{grammar}
[(colon){$\rightarrow$}]
[(semicolon)$|$]
[(comma){}]
[(period){\\}]
[(quote){\begin{bf}}{\end{bf}}]
[(nonterminal){$\langle$}{$\rangle$}]

<net>:["\bf pure"],"\bf net",<vertex-name>,"[",<config-params>,"]",\\
	"\tt\bf (",<in-chans>,"\tt |",<out-chans>,"\tt\bf )", <decls>, "\bf connect", <wiring>, "\bf end".

<decls>: [<decl>]*.
<decl>: <net>;<synchroniser>;<morphism>.

<morphism>: "morph","(",<size>,")","\{"<morph-list>,["where",<override-list>]"\}".
<morph-list>:<morph>,[","<morph>]*.
<size>:<id>.

<morph>:<split>,"/"<map-list>,"/",<join>\\
	;<split-map-list>,"/",<join>\\
	;<split>,"/",<map-join-list>.

<map-list>:<map>,[","<map>]*.
<split-map-list>:"(",<split-map>,[","<split-map>]*,")".
<map-join-list>:"(",<map-join>,[","<map-join>]*,")".

<map>:[<int>,":"],<id>.
<split-map>:<split>,"/",<map>.
<map-join>:<map>,"/",<join>.
<split>:<id>.
<join>:"O'",<id>;"S'",<id>;"U'",<id>.

<override-list>:<override>,[",",<override>]*.
<override>:<join>,"{\tt\bf ..}",<split>,"=",<synch>.
<synch>:<id>.

\end{grammar}
\end{framed}
\caption{General syntax of the {\bf net} environment\label{fig:net}}
\end{figure}

The construction of networks in \ak\ is hierarchical: vertices are combined into a subnetwork, which in turn can act as a vertex in a larger network, etc. This is achieved with the help of a construct called \verb$net$. The general syntax of the net declaration is as follows in fig \ref{fig:net}.
Here {\it vertex-name} is the name of the resultant vertex, {\it config-params} are the configuration parameters of the net, which are fully analogous to those of the synchroniser, as are the channel lists.
The construct {\it decls} declares all compound vertices in it using nested \verb$net$ constructs, and also any synchronisers and morphisms (see below) 
and finally the {\it wiring} construct wires up the subgraph that represents the subnetwork using the facilities
described in section \ref{sec:wiring}, any synchronisers whose definitions are visible in the scope of the current \verb$net$ environment, 
and boxes referred to by name. 

The vertices in the subgraph after the wiring should only have output channels mentioned in the \verb$net$ header. Any extra input channels are, nevertheless, allowed, and will, as is common elsewhere, be plugged up with $\sigma_0$. Also the output channels mentioned in the header but not supported by the wiring inside the connect clause will be plugged up likewise. The \verb$net$ construct declares a single compound vertex with channels precisely as specified in the header.

\paragraph{Pure nets.} The vertex declared by the \verb$net$ construct can be made to behave exactly like a box. This is achieved by using the \verb$pure$ keyword
and supplying the appropriate category (see section \ref{sec:category}) as a prefix to the name separated by semicolon. For example: \verb$du:foo$ qualifies
the pure net being defined as a dyadic unordered reductor. The category may be omitted together with the semicolon in which case the pure net is assumed to be a
transductor.

The numbers of input and output channels should correspond to the chosen
category and the depth of all channels must be assumed to be 0. Inductors yield a $\sigma_0$ on its output channels to mark the end of the sequence,
and the reductor determines the end of an input sequence when it encounters a $\sigma_0$ at its input channel. This way if a true box is placed in a pure \verb$net$
environment it will operate correctly and so will be a complex network if it obeys these rules. The environment itself will correctly transfer segmentation marks other than $\sigma_0$ from input to output with the appropriate adjustment; the network inside will always see all input and output channels as being depth 0.

A pure net may have synchronisers inside. The \verb$net$ environment will interfere with their operation in the following way:
\begin{itemize}
\item for the transductor category: upon the yield of a message on all output channels the environment will abort the execution of all boxes that are still running,
flash all channels and force all internal synchronisers back to the start state (which will have the same effect as a proper transition to that state without sending anything). If quiescence of the internal network is detected prior to that, i.e. all boxes are inactive and all synchronisers are expecting messages on empty
channels, they are brought back to the start state and all the internal channels similarly flashed.
\item for the inductor category: the same happens after the inductor has yielded a $\sigma_0$.
\item for the reductor category: the same happens after the reductor has encountered a $\sigma_0$ at its input and has sent its outut message.
\end{itemize}

Recall that TPL vertices are boxes, synchronisers and finally, nets as defined by the construct in question. The $\langle$decls$\rangle$ clause of the {\bf net}
environment must declare the nets and synchronisers that are to be used in the wiring expression, the latter following the syntax already defined in section \ref{sec:synch}. The boxes do not need to be declared since their only attribute
in TPL is category, which is specified as part of the box name precisely in the manner that has just been described for pure nets. 
What does require a declaration, however, is the algebraic properties
of boxes that will be introduced in the next section.

\begin{figure}
\includegraphics[scale=0.75]{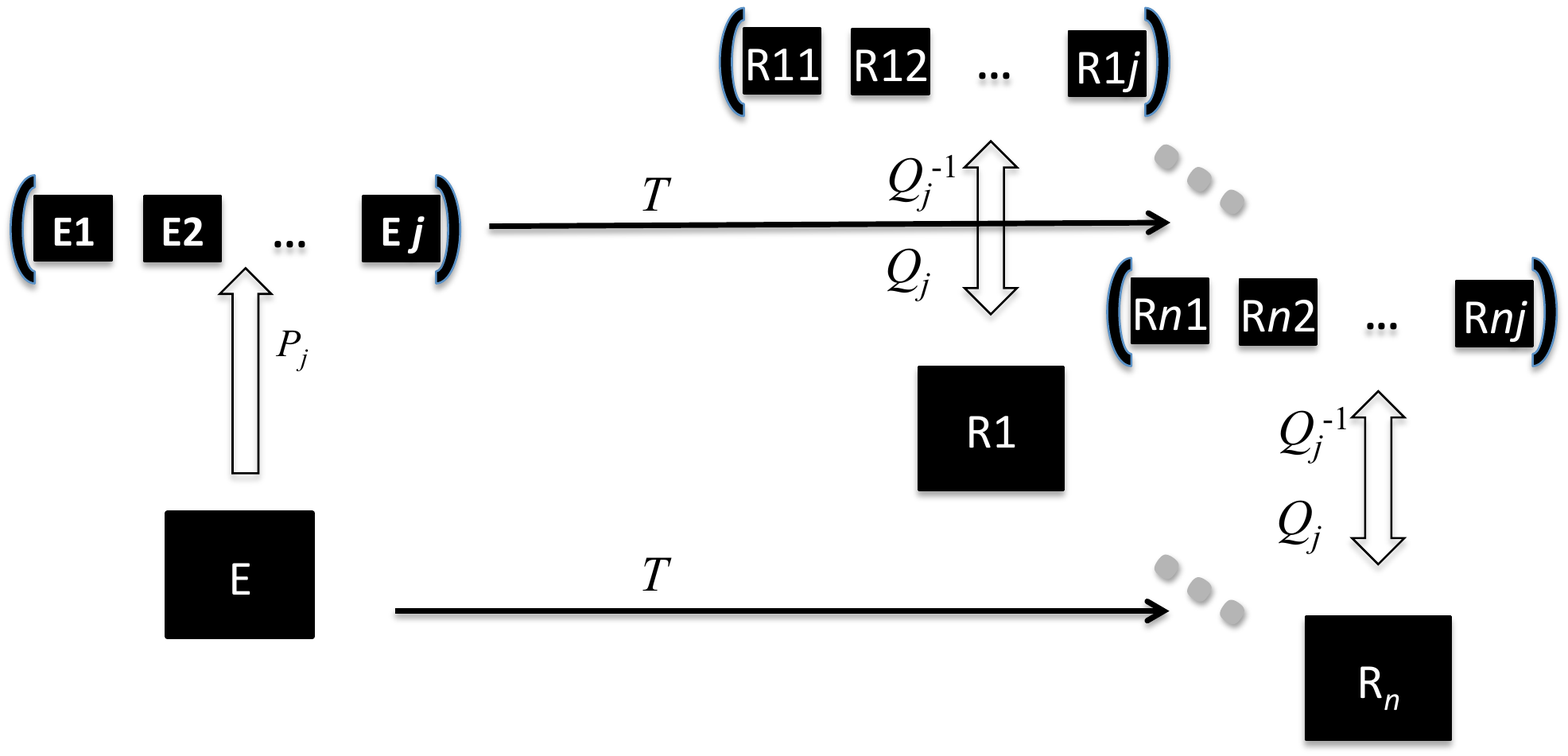}\\
\vspace{-2.5in}
\caption{Loxomorphism of boxes
\label{fig:loxo}}
\end{figure}
\subsection{Morphisms}
TPL uses pressure as a trigger for proliferation and this adaptation mechanism can be effective in a situation in which a box has a large supply of messages
to work on and when the demand for result messages is also high. There is, however and important class of situation where the supply and demand are of order unity,
but a transductor box admits a divide-and-concur strategy, which we will describe formally by giving the following definitions.
\begin{mydef}
A transductor $n${\rm T} is {\it eager} if it responds to any input message with exactly one message on each of its $n$ output channels.
\end{mydef}
\begin{mydef}[Loxomorphism]
For integers $K$, $n$ and the monadic reductor category $\alpha\in\{${\rm O,S,U}$\}$, a loxomorhism $\Lambda_{Kn}^\alpha$ is a triplet $(P_j, T, Q_j^{-1})$, where
$T$ is an eager $n${\rm T} transductor and
for all $j\le K$, $P_j$ and $Q^{-1}_j$, are {\rm 1I} inductors, $Q_j$ is a {\rm 1M$\alpha$} reductor such that the network $Q^{-1}_j${\tt .\kern-3pt.}$Q_j$ behaves
as a straight channel,  and the diagram in fig~\ref{fig:loxo} is commutative:
\[
P_j\hbox{\tt .\kern-3pt.}T = T\hbox{\tt .\kern-3pt.}\hbox{\Large\tt|\kern-5pt|}_{i=1}^n \langle\verb$_$i\mid Q^{-1}_j\mid \verb$_$i\rangle
\]
\end{mydef}

The term loxomorphism, introduced here, is based on the Greek prefix loxo-, which means ``oblique'' and
which loosely refers to the fact that the diagram is lopsided:
the arrow $P_j$ is single and not necessarily reversible,
while the right one $Q^{-1}_j/Q_j$ is generally multiple and must be reversible.
The practical significance of loxomorphisms is that they permit splitting an input message E into a series of, expectedly, smaller messages, E1, E2, $\ldots$, E$j$, 
each requiring less time
to process, and which can be processed in parallel. The results form a series of messages as well, which are an image of the
intended single result under a transformation similar (but not necessarily identical)  to that which generate the first series. According to the definition, a reduction
must exist that transforms the result series back to the single result. Notice that the figure suggests that the inductor $P_j$ produces exactly $j$ messages
in the series, but that is not strictly required. The run time system will expect the actual number of messages to be a monotonic function of $j$, just as it will expect
the processing costs associated with a single application of $T$ to be monotonically decreasing with $j$ for a given message E. The process or replacing
the original transductor by an inductor-transductor-reductor pipeline is called {\em fragmentation} and is triggered automatically in the same way as the
process of proliferation is activated. 

\subsection{Declarations of morphisms}

The syntax of the morphism declaration is shown in fig \ref{fig:net}. The nonterminal {\it size} defines the name of the variable that should be present
in the input message in order for the morphism to be applicable. The splitter uses its integer value to determine how many submessages the original message
should be split into. When fragmentation occurs, the run time system chooses the value of {\it size} and extends the input message with the variable specified 
as {\em size} giving it a certain value.   

Loxomorphisms can be grouped to share the {\em size} variable and the inductor $P$ or  the reductor $Q$ (see the definition) in order to share the splitting or joining mechanism among different types of partitioned processing. Consequently the syntax provides two forms of bracketing for associating a single inductor or 
reductor with several transductors. Also sharing both the $P$ and the $Q$ among transductors is supported. Each transductor is named in the {\it map} clause
and is optionally specified with the number of output channels as an integer prefix separated by a colon. This corresponds to the $n$ parameter 
in $\Lambda_{Kn}^\alpha$. 
The {\em join} clause names the reductor and specifies
its concurrency category: ordered, segmented or unordered, which is the $\alpha$ parameter in $\Lambda_{Kn}^\alpha$. 
Notice that ordered reductors are allowed in morphisms despite the fact that they can only be executed
sequentially: their computational cost could be much smaller than that of the transductor in the morphism group, and that will be taken into account by the DIL
observation loop to be discussed elsewhere. 

A very important mechanism that makes loxomorphisms efficient in \ak\ is the pipelined override introduced by the {\bf where} clause. It is often the case
that two or more transductors are chained with a serial connection in order to work consecutively on the same message by applying different kinds of processing to it.
Under appropriate pressure conditions fragmentation is likely to occur in the whole chain, since each stage of the pipeline conserves the number of messages. 
If all stages are fragmented using appropriate morphisms, the joiners and splitters are inserted between transductors to gather the results into one message and split it back out again. When the joining and the splitting possess some degree of spatial locality, the creation of a single message only to split it again is unnecessary. Instead, one or more fragments of the result of one stage can be combined straight into a fragment required for the partitioned processing of the next. Consequently, a  (table) synchroniser specifically written for the purpose can have the same effect as the serial combination of a joiner and a splitter. The morphism declaration allows the
coordination programer to indicate that such synchronisers are available and should be used whenever a particular joiner/splitter pipeline is inserted by the
run-time system as a result of fragmentation.         

Finally, a few comments about the data relations introduced by the morphisms. A message coming to a fragmentable transductor must be typed to indicate 
how many fragments it can be split into, which corresponds to the value of $K$ in $\Lambda_{Kn}^\alpha$. 
For instance a 1d array of size $N$ cannot be split into more than $N$ fragments and that is something that the run-time
system needs to know to be able to correctly choose the value of the {\em size} parameter. Consequently a morphism declaration generates a type constraint
of the kind that synchronisers do, and which must be satisfied by the environment. These issues are tackled in the next chapter where the constraint system 
of \ak\ is presented.

\chapter{Constraint Aggregation Layer}

\section{Philosophy}

The Constraint Aggregation Layer of \ak\ is an analysis layer above the TPL. Recall that the TPL describes the behaviour (i.e., "progress") of vertices in an \ak\ network under the streams of incoming and outgoing messages: the vertices' activation, proliferation, fragmentation, suspension, etc., as well as the topology of the streaming network itself. In so doing, the TPL almost entirely abstracts from the content of the messages: they are seen as completely opaque and are only subdivided into $\sigma$s and data messages, the former being self-contained and the latter indivisible and incomprehensible to TPL. The only exception to this is the synchroniser: it is able to peek into the incoming messages and form new messages on the output by engaging the (external) mechanisms of coercion and concatenation. Even there the invasion of the message privacy is minimal: a synchroniser is empowered to determine the variant of, and/or extract integers from an input message, and to combine messages and/or integers into a new message. TPL does not have the knowledge of message structures to accomplish those by itself; it delegates such analyses to CAL by stating constraints that define the relationship between input and output message formats. CAL resolves the constraints in the context of the entire network and makes the corresponding data-manipulations sufficiently well defined for code generation.

\paragraph{Passports.}In data communication --- be it the sending of a message from one place to another, or,  more generally,  of messages from a group of places for copying to a group of recipients --- the static correctness of the channel demands that the statically guaranteed properties of an output message be sufficient to satisfy the static requirements of its recipients. For instance, a producer of messages that are consumed by a box that takes a square root of a real number must ensure that any messages are numbers, and that the numbers are real and nonnegative. Generally speaking, a message is a collection of items that can be distinguished (either positionally or by name), and so a message is endowed with a collection of item properties, and its recipient with a collection of item constraints. The guaranteed properties of the message can be ascribed to its originator's output channel just as the input requirements are the properties of the recipient's input channel. Since, generally speaking, a vertex is both an originator of its output messages and the recipient of the input ones, and since the input constraints are the necessary condition for the vertex to operate correctly and hence to guarantee the output properties, the vertex can be abstracted with respect to its data-transformation behaviour as an implicative statement $p\Rightarrow P$.  Here $p$ is the conjunction of all the requirements and $P$ is the conjunction of all the guarantees. We will call these implications box {\em passports}.

\paragraph{Terms.} A channel connecting two vertices will thus connect two sets of properties: the requirements and the guarantees. Since the sender's guarantees must be sufficient to satisfy the recipient's requirements, a channel can also be conceptualised as an implication. However, unlike the vertices, the channel does not {\em change} the messages that are communicated on it. Moreover, the {\em only} thing that is known about a message is the conjunction of its sender's guarantees. Consequently the channel implication is equal in force to the subset relation between the set that the messages of the originator vertex are guaranteed to fall within and the recipient's set of acceptable input messages. Such relation can be expressed in symbolic form in a term algebra, where the variety of output messages is represented by a term and the input requirements also by a term. Variables can occur in terms, which stand for (yet unknown) subterms. The input and the output parts of the passport may share some of the variables, which is how the relation between input and output properties of a vertex is established. This happens when the vertex is polymorphic with respect to its input types. In contrast to the conventional typing of functions, an \ak\ vertex can be (non-parametrically) polymorphic to the output types alone as well, not necessarily in conjunction with input type polymorphism .

\paragraph{Putting it all together.} This is how the TPL and CAL work together. Each box is now represented as a combination of a source code and a triad of the box name, box category and  CAL passport. The passport generally uses term variables to represent requirements and guarantees. Those variables can also be used in the box source as external parameters for macros and/or compilation parameters accessible by the compiler. Such use may reflect either genericity of the component or its parametric or nonparametric polymorphism. Another significance of term variables may be for the purposes of self-tuning: choosing the correct platform parameters, implementation scheme or even switching between different algorithms that compute the same results with an efficiency dependent on the properties represented by the term variables. An \ak\ compiler performs its first (Constraint Aggregation or CA) pass by only taking the above-mentioned triad and the coordination program written in \ak/TPL. During the CA pass, the topology of the network is extracted from the \ak\ program, the properties of the synchronisers with respect to input terms (i.e. their ``passports'' if they had one) are inferred from each synchroniser program, and the process of juxtaposition and constraint solving is performed to instantiate all term variables. As a side effect, a proof is obtained that the constraint system is satisfiable, which indicates that all components have received sufficient assurances to guarantee their output, and consequently the whole program that will be generated next is consistent and type correct. That concludes the CA pass.

The second, and final,  pass of the \ak\ compiler breaks down into two independent stages. First of all, with all the term variables instantiated the box compiler(s) are now able to complete the tuning stage, generate any macro expansions that use the term variables and finally produce binary code for whatever platform the variables may tell it (them) that the code 
is required. At the same time, any message manipulation code requested by the synchronisers is produced by the 
box-language code generators. The second stage is simultaneously performed by the \ak\ compiler and includes 
generation of the binary code from the \ak/TPL program and linking it with the box code received from the first stage.

Note that the combination of CAL and TPL still leaves open any issues of data management (as opposed to message management). Messages only contain constants or constant references to data objects. The objects themselves have to be created, accessed, including concurrent partitive access with modifications possibly under transaction control. Objects may need to be moved around the platform and disposed of. All these actions may depend on the statistics of object access, requiring instrumentation and feedback. Also data management has to be made available to boxes in the form of API. All of these issues are the prerogative of the Data and Instrumentation Layer (DIL) covered in Chapter 3. One issue that is worth mentioning already in the current chapter is that the DIL will require knowledge of the data {\em kind} in order to be able to create/initialise/delete objects, and so CAL properties will be required for its correct functioning. Consequently the DIL is positioned on top of the CAL.

By now the reader will understand that the role of the CAL in \ak\ is similar to the role of the type system in a conventional (non-coordination) programming language. The CAL is, in a way, a universal type system in the sense that it does not fix the structure and meaning of the type assertions that boxes may choose to import and export. It instead provides a constraint programming framework in which a wide variety of assertions can be formulated. It relies on general-purpose constraint solving as a means of type checking, type inference and most general subtyping.

In the sequel we introduce the syntax of the CAL passport and discuss the basic aggregation mechanisms.

\section{Specifying terms}

CAL is based on the Message Definition Language (MDL) which is a language of abstract terms that are built 
recursively from the ground up. Structurally they are symbolic trees with the following kinds of leaf:

\def\true{\hbox{\bf true}}
\def\false{\hbox{\bf false}}

\begin{description}
\item[symbol] is an identifier representing a certain finite quality, such as {\em int}, {\em even}, {\em char}, {\em red}. They are the main building block of the term structure. A symbol may not start with a dollar sign.
\item[number] CAL terms can use numbers in various forms. C conventions apply here for integer (signed and unsigned), fractions including floating-point, and bit masks in hexadecimal form.
\item[string] ASCII strings in double quotes also follow the C convention
\item[variable] Term variables are represented by identifiers that start with a dollar sign. They range over terms.
\item[flag]	A flag is a Boolean variable that ranges over a set of two symbols: $\{\true,\false\}$. Syntactically a flag is indistinguishable from a symbol, but it only occurs in certain contexts and is unumbiguously identified by them.
\end{description}

Symbols, numbers and strings are considered mutually distinct, while variables can be tied with each other or a variable can be tied to a term. Terms are built recursively using the following types of constructor(see fig~\ref{fig:mdl}):
\def\dbbr{\tt\bf\textbar\kern-1pt\textbar}
\begin{figure}[htb]
\begin{framed}
\small
\begin{grammar}
[(colon){$\rightarrow$}]
[(semicolon)$|$]
[(comma){}]
[(period){\\}]
[(quote){\begin{bf}}{\end{bf}}]
[(nonterminal){$\langle$}{$\rangle$}]

<term>:<symbol>;<number>;<string>;<variable>;\\
<tuple>;<list>;<record>;<choice>;<switch>.

<tuple>:"(",<term>,[<term>]*,")".
<list>:"["<term>[",",<term>]*,["\dbbr",","<list-tail>],"]".
<list-tail>:<list>;<variable>;"\bf nil".

<record>: "\{",<lab-guarded-list>,[<record-tail>],"\}".
<choice>:  "(:",<lab-guarded-list>,[<choice-tail>],":)".
<lab-guarded-list>:<member>[",",<member>].
<member>:<label>,["(",<guard-exp>,")"],":",<term>.
<label>:<symbol>.
<record-tail>:<record>;<variable>;"\bf nil".
<choice-tail>:<choice>;<variable>;"\bf none".

<switch>:"\la",<guarded-member>,[","<guarded-member>]*,"\ra".
<guarded-member>:<guard-exp>,":",<term>.

<guard-exp>:\true;\false;<flag>;<not>;<or>;<and>.
<flag>:<symbol>.
<not>:"(\bf not ",<guard-ex>,")".
<and>:"(\bf and ",<bool-list>,")".
<or>:"(\bf or ",<bool-list>,")".
<bool-list>:<guard-exp>,[<guard-exp>]*.
\end{grammar}
\end{framed}
\caption{MDL syntax\label{fig:mdl}}
\end{figure}

\begin{description}
\item[tuple] A tuple term is a collection of terms in linear order. Each term is identified by its position in the tuple.  A tuple node is written as a parenthesised space-separated list of terms, e.g.: \verb$(a 24 b "qq")$, \verb$(alpha b (c d))$, etc. A one-tuple is equivalent to its member written without the parentheses, e.g. \verb"((x y))" is the same as \verb"(x y)".

\item[list] A list is an extensible collection of terms in linear order. Syntactically a list is a pair $h,t$, where $h$ is any term
and $t$ is a list term or \verb$nil$, the latter symbol representing the empty list, written as $[h\| t]$. In order to avoid a large number of enclosed parenthesis, the following syntactic sugar is supported:
\[
[e_1\| [e_2 \| t]] = [e_1,e_2\| t]\,,
\]
and \verb$nil$ as $t$ can be omitted together with the preceding bar. Also, for order-theoretical reasons that will be clear later, empty lists cannot be nested: \verb$[nil]$=\verb$nil$.

\item[record] A record is a term representing
a collection of label-term pairs. The label of a pair is an arbitrary symbol except no
two pairs can have the same label. Records are comma-separated lists enclosed in braces and written
in tail form: \verb${$$\{l\hbox{:}t || s\}$, where $l$ is a label,
$t$ is the term associated with the label and $s$ is the rest of the the record.
If $s$ is empty (also represented in MDL as  \verb$nil$) then the bar and $s$ can be omitted. As with lists, multiple enclosed braces can be fused into one, e.g.  \verb"{a:x,b:y||{...}}" is a shorthand for \verb"{a:x||{b:y||{...}}}"

Let $\mathcal L$ be the set of labels occurring anywhere in the passports of a program.
For a record $R$, define  $\hat{R}\subseteq {\mathcal L}$ as the label set of R. A record $\{l:t || R\}$
is well-formed if and only if  $l\not\in\hat{R}$.
CAL records have the following basic property: for any record R,
$l_1\ne l_2\in {\mathcal L}$, $l_{1,2}\not\in\hat{R}$
and any $t_{1,2}$
\[
\{l_1\hbox{:}t_1,l_2\hbox{:}t_2|| R\}=\{l_2\hbox{:}t_2,l_1\hbox{:}t_1|| R\}
\]
Record members can be present in the record conditionally. This is achieved by using a guard
term after the label as follows: $\{l(g)\hbox{:}t||R\}$. The record contains the label $l$ and its
associated term $t$ if and only if the guard $g$ is \true, otherwise the record
is identical to $R$. A label may have several conditional occurrences in the same record term; however, the term is only well-formed if there exists no more than one occurrence of any label in it with the \true{} guard.  An unguarded member of a record is assumed to have \true{} as the guard.

\item[choice] A choice term is intended to represent a collection of {\em alternative} terms.
Choices are comma-separated lists enclosed in colonised parentheses: \verb"(:" and \verb":)", e.g.
\verb"(: a:t1, b:t2 || none :)".
Like records choices are written in tail form with the same shorthand convention except the symbol \verb$none$ is used rather than the symbol \verb$nil$; the former represents the vacuous choice
that does not match any data format. The same equational property exists for
choices as the one for records, and choice members (also called {\em alternatives}) can be
guarded. An unguarded choice is equivalent to a similar guarded choice where all the guards being \true{}. Several guarded occurrences of the same label are allowed, and the choice is well formed when no more than one guard is positive for any given label.

\item[switch] A switch is a collection of guarded terms representing exactly one of them depending on the value of the 
guards. Syntactically, a switch is a comma-separated list of guarded terms enclosed in angular brackets. For a switch
to be well formed exactly one guard must be \true, e.g. \verb"< a:[1,2], b:[3]>" is a well-formed switch provided 
that the flags $a$ and $b$ satisfy  $a\wedge \neg b=\true$, and it is equal to \verb$[1,2]$ if $a$ is positive and \verb$[3]$ 
otherwise.

\end{description}

\paragraph{Boolean logic.} For the purposes of supporting complex guard terms, and more generally, for building conditional terms,  we introduce Boolean expressions that are, syntactically, trees of tuples with flags as leaves. Figure~\ref{fig:mdl}
defines the syntax of the Boolean expression in a straightforward fashion using standard Boolean functions. Note that 
since Boolean expressions only occur in special positions unambiguously defined as such by the syntax, flags need 
not be syntactically differentiated from symbols: in a Boolean context any symbol other than \true and \false denotes
a Boolean-valued flag. That is in contrast to term-valued variables, which have to be lexically different from symbols
as the former occur in the same context as the latter.

\section{Relations on terms}

Consider MDL terms without variables and flags. Note that such terms can be assumed not to contain switches without
loss of generality. In any records or choices within a term all labels can also be assumed to be pairwise distinct, and all guards to be \true. Such terms are said to be {\em ground}.

We now define a {\em seniority relation} on ground terms:

\begin{mydef}
A terms $t_2$ is said to be {\em\bf senior} to a term $t_1$, $t_1 \sqsubseteq t2$, iff at least one of the following six conditions is satisfied.
\begin{enumerate}

\item $t_2=$\verb"nil".

\item $t_1$ and $t_2$ are the same symbol, number or string.

\item $t_1=[t^1_1 ... t^k_1]$ and $t_2=[t^1_2 ... t^m_2]$ for some $k\ge m$ and
$t_1^i$, $t_2^i$ provided that for all $i=1..m$ $t^i_1\sqsubseteq t^i_2$.

\item $t_1=(t^1_1 ... t^k_1)$ and $t_2=(t^1_2 ... t^k_2)$ for some $k\ge1$ and
$t_1^i$, $t_2^i$ provided that for all $i=1..k$ $t^i_1\sqsubseteq t^i_2$.

\item $t_1=\{l^1_1\hbox{\rm:}t^1_1,..,l^k_1\hbox{\rm:}t^k_1\}$ and $t_2=\{l^1_2\hbox{\rm:}t^1_2,..,l^m_2\hbox{\rm:}t^m_2\}$, where $k\ge m$ provided that
\[(\forall j\le m, \exists i\le k) l^i_1=l^j_2\; \hbox{\rm\bf and}\; t^i_1\sqsubseteq t^j_2.\]

\item $t_1=\hbox{\rm (:}\,l^1_1\hbox{\rm:}t^1_1,..,l^k_1\hbox{\rm:}t^k_1\,\hbox{\rm :)}$
and $t_2=\hbox{\rm (:}\,l^1_2\hbox{\rm:}t^1_2,..,l^m_2\hbox{\rm:}t^m_2\,\hbox{\rm :)}$,
where $k\le m$, provided that
\[(\forall i\le k, \exists j\le m) l^i_1=l^j_2 \; \hbox{\rm\bf and}\; t^i_1\sqsubseteq t^j_2.\]

\end{enumerate}

If $t_1\not\sqsubseteq t_2$ and $t_2\not\sqsubseteq t_1$ the terms are said to be incommensurable.
If $t_1\sqsubseteq t_2$ and $t_2\sqsubseteq t_1$ we write $t_1=t_2$ call them equal.
If $t_1\sqsubseteq t_2$ and $t_1\ne t_2$ then $t_2$ is said to be {\em properly senior} to $t_1$.
Finally, if $t_2$ is (properly) senior to $t_1$ then $t_1$ is (properly) junior to $t_2$. 
\end{mydef}

As follows from the definition the seniority of records is established by the principle that a larger record that contains a given one as a subset of its members is junior to it, and a larger choice of the same kind is senior. Tuples and atoms must have the same arity and the members of one tuple are required to be senior to the other (as are the corresponding members of records and choices as well).

\newtheorem{prop}{Proposition}
\begin{prop}
The seniority relation $\sqsubseteq$ defined above is a partial order.
\end{prop}
It can be verified directly from the definition by structural induction that the seniority relation is reflexive, transitive and antisymmetric, which proves the proposition.

The seniority relation defines what kinds of messages (described by a term) can be regarded as a particular case of a given message variety (described by another term) in the sense
of carrying sufficient information to be properly converted\footnote{Here we consciously avoid
the word ``type'' since the interpretation of MDL is fully abstract while box languages may have
a very specific idea of what types are, but it would not be wrong to think of terms as generalised ``types''.}.
For example, a record
with more members can easily be transformed into a smaller record by leaving some members out. A choice can be converted into a larger choice by never using the variants of the larger choice whilst (notionally) stating that they are included.
The extreme case of a message is one that does not carry data at all. Clearly any message can be converted to this case, which represents the top element of the partial order. We will denote it as \verb"nil". An empty choice differs from the empty record in that the latter can be used as a trigger for some computation in a synchroniser or even a box, while the empty choice when used to describe received data signifies that no data of any kind can ever be received. The empty choice is equal to 
the symbol \verb$none$ and represents the bottom of the choice order.

\begin{prop}
The join of any two records, $P=\{l^1_1\hbox{\rm :}t_1^1,..,l^k_1\hbox{\rm :}t^k_1\}$
and $Q=\{l^1_2\hbox{\rm :}t^1_2,..,l^m_2\hbox{\rm :}t^m_2\}$ exists and is computed
as follows. First assume that the labels in $P$ and $Q$ are enumerated in such a way that
the first $r\ge0$ labels occur in both $P$ and $Q$. If $r=0$ then $P\sqcup Q=$\verb"nil".
Otherwise
\[
P\sqcup Q=\{(l^1_1=l^1_2):t^1_1\sqcup t^1_2,..,(l^r_1=l^r_2):t^r_1\sqcup t^r_2\}\,.
\]

\end{prop}

The proof follows from the observation that any upper bound is a record; that it can include at most (so as to be minimised) all the common labels of $P$ and $Q$; and that the terms associated with those labels can only be senior to the terms shown above.

\begin{prop}
The join of any two choices $A=\hbox{\rm (:}\,l^1_1\hbox{\rm :}t_1^1,..,l^k_1\hbox{\rm :}t^k_1\,\hbox{\rm :)}$ and
$B=\hbox{\rm (:}\,l^1_2\hbox{\rm :}t^1_2,..,l^m_2\hbox{\rm :}t^m_2\,\hbox{\rm :)}$ exists and is computed as follows.
First assume the first $r\ge0$ labels in $A$ and $B$ are common.
Then the join $A\sqcup B$ is the following term:
\[
A\sqcup B = \hbox{\rm (:}\,l^1_1\hbox{\rm :}t_1^1,..,l^{k}_1\hbox{\rm :}t^{k}_1, l^1_2\hbox{\rm :}t^1_2,..,l^{m}_2\hbox{\rm :}t^{m}_2\,\hbox{\rm :)}
\]
if $r=0$, and
\[
A\sqcup B = \hbox{\rm (:}\,
(l^{1}_1=l^1_2)\hbox{\rm :}t^{1}_1\sqcup t^{1}_2
,..,(l^{r}_1=l^{r}_2)\hbox{\rm :}t^{r}_1\sqcup t^{r}_2,
l^{r+1}_1\hbox{\rm :}t^{r+1}_1,..,l^{k}_1\hbox{\rm :}t^{k}_1,
l^{r+1}_2\hbox{\rm :}t^{r+1}_2,..,l^{m}_2\hbox{\rm :}t^{m}_2
 \,\hbox{\rm :)}
\]
otherwise.
\end{prop}
The proof follows from the observation that any upper bound will be a choice; that it will include at least (so as to be minimised) all the labels from both $A$ and $B$; and that the terms associated with the common labels must be senior to those in $A$ and $B$.

\begin{prop}
For two list terms $t_1=[h_1\|t_1]$ and $t_2=[h_2\|t_2]$,  $t_1\sqcup t_2 =[h_1\sqcup h_2 \| t_1\sqcup t_2]$.
\end{prop}
The proof is straightforward. Note that that the join of two lists of mutually incommensurable elements is \verb$nil$,
due to the fact, mentioned earlier, that empty lists do not nest.  

\begin{prop}
For two terms $t_1$ and $t_2$ that are any of the following: symbol, number or string,
$t_1\sqcup t_2 = t$ if $t_1=t_2=t$ and \verb"nil" otherwise. For two tuples the join equals
\verb"nil" unless they have the same arity:
\[
(t^1_1\,...\,t^k_1) \sqcup (t^1_2\,...\,t^k_2) = (t^1_1\sqcup t^1_2,..t^k_1\sqcup t^k_2)\,.
\]
\end{prop}
The proof directly follows from the definition of the seniority relation on terms.

\begin{prop}
The seniority relation makes the set of all ground MDL terms a join-semilattice.
\end{prop}
This is a consequence of propositions 2--5 and the fact that terms of different nature are junior only to \verb$nil$.

\begin{prop}
The meet of two records $A=\{l^1_1\hbox{\rm :}t_1^1,..,l^k_1\hbox{\rm :}t^k_1\}$ and
$B=\{l^1_2\hbox{\rm :}t^1_2,..,l^m_2\hbox{\rm :}t^m_2\}$ may or may not exist
and is computed as follows.
First assume the first $r\ge0$ labels in $A$ and $B$ are common.
If $r=0$ then the meet $A\sqcap B$ exists and is given by the following term:
\[
A\sqcap B = \{l^1_1\hbox{\rm :}t_1^1,..,l^{k}_1\hbox{\rm :}t^{k}_1, l^1_2\hbox{\rm :}t^1_2,..,l^{m}_2\hbox{\rm :}t^{m}_2\}\,.
\]
Otherwise the meet exists if and only if all $t^{1}_1\sqcap t^{1}_2,..,t^{r}_1\sqcap t^{r}_2$ exist. In such a case
\[
A\sqcap B = \{
(l^{1}_1=l^1_2)\hbox{\rm :}t^{1}_1\sqcap t^{1}_2
,..,(l^{r}_1=l^{r}_2)\hbox{\rm :}t^{r}_1\sqcap t^{r}_2,
l^{r+1}_1\hbox{\rm :}t^{r+1}_1,..,l^{k}_1\hbox{\rm :}t^{k}_1,
l^{r+1}_2\hbox{\rm :}t^{r+1}_2,..,l^{m}_2\hbox{\rm :}t^{m}_2
 \}\,.
\]
\end{prop}
The proof follows from the observation that any lower bound will be a record; that it will necessarily include all the labels from both $A$ and $B$; and that the terms associated with the labels must be
junior to the ones used above. The meets of the element terms may not exist, when for example,
the terms are two different symbols or numbers: there does not exist a single symbol or number that is junior to two different ones at the same time.

\begin{prop}
The meet of two choices, $P=\hbox{\rm (:}\,l^1_1\hbox{\rm :}t_1^1,..,l^k_1\hbox{\rm :}t^k_1\,\hbox{\rm :)}$
and $Q=\hbox{\rm (:}\,l^1_2\hbox{\rm :}t^1_2,..,l^m_2\hbox{\rm :}t^m_2\,\hbox{\rm :)}$ may or may not exist and is computed as follows.
First assume that the labels in $P$ and $Q$ are enumerated in such a way that
the first $r\ge0$ labels occur in both $P$ and $Q$. If $r=0$ then the  meet exists only when
$P=Q=$\verb"none", in which case $P\sqcap Q=$\verb"none" also.

Otherwise the meet exists if and only if all $t^{1}_1\sqcap t^{1}_2,..,t^{r}_1\sqcap t^{r}_2$ exist. In such a case
\[
P\sqcap Q=\hbox{\rm (:}\,(l^1_1=l^1_2):t^1_1\sqcap t^1_2,..,(l^r_1=l^r_2):t^r_1\sqcap t^r_2\,\hbox{\rm :)}\,.
\]

\end{prop}

The proof follows from the observation that any lower bound is a choice; that it must include at most
(so as to be maximised) all the common labels of $P$ and $Q$; and that the terms associated with those labels can only be junior to the terms shown above.

\begin{prop}
The meet of two nonempty lists  $t_1=[h_1\|t_1]$ and $t_2=[h_2\|t_2]$ is given by  
$t_1\sqcap t_2 =[h_1\sqcap h_2 \| t_1\sqcap t_2]$, and if $t_2=$\verb$nil$, $t_1\sqcap t_2 =t_1$.
\end{prop}

\begin{prop}
For two terms $t_1$ and $t_2$ that are any of the following: symbol, number or string $t_1\sqcap t_2$
exists if and only if $t_1=t_2=t$, in which case $t_1\sqcap t_2 = t$ also. For two tuples, the meet only 
exists when they have the same arity $k$:
\[
(t^1_1\,...\,t^k_1) \sqcap (t^1_2\,...\,t^k_2) = (t^1_1\sqcap t^1_2\,...\,t^k_1\sqcap t^k_2)
\]
provided that all $t^1_1\sqcap t^1_2,..,t^k_1\sqcap t^k_2$ also exist.
\end{prop}
The proofs follow from the definition of the seniority relation on terms.

\subsection{Passport syntax}

The passport syntax is given in figure \ref{pass-syntax}. A parenthesised list of input terms is followed by the constraint section and then by a similar list of output terms. The constraint section is intended for constraints of various nature. In the current version of the document only order-theoretic constraints are defined. They consist of the left hand side that can be either an individual term or a join, a relation sign, which can be either equality $=$ or seniority $\sqsubseteq$ coded as \verb"<=", and the right hand side that can be either a term or a meet (the join and meet, respectively, are not specially marked; they are assumed in the semicolon-separated lists of terms.

\begin{figure}
\begin{framed}
\begin{grammar}
[(colon){$\rightarrow$}]
[(semicolon)$|$]
[(comma){}]
[(period){\\}]
[(quote){\begin{bf}}{\end{bf}}]
[(nonterminal){$\langle$}{$\rangle$}]
<passport>:"vertex",<name>,<body>.
<name>:<id>.

<body>: \verb"<",[<input>,[",",<input>]*],\verb"|",<constraints>,\verb"|",[<output>,[",",<output>]*],\verb">".

<input>:<term>.
<output>:<term>.

<constraints>:[<lhs>,<rel>,<rhs>,"."]*.
<lhs>:<term>,[";",<term>]*.
<rhs>:<term>,[";",<term>]*.
<rel>:\verb" =";"{\tt <=}".

\end{grammar}
\end{framed}
\caption{The CAL passport\label{pass-syntax}}
\end{figure}

\subsection{Synchroniser passport}

As mentioned before, synchronisers rely on both TPL and CAL for their definition. The CAL aspects of a synchroniser are confined to the CAL terms for its input and output channels. Those terms are not straightforward since they have to be fairly generic to match the broadest possible formats of producer and consumer messages involved in the act of synchronisation.  On the other hand, the synchroniser passport is produced solely on the basis of the synchroniser code, exclusively by program analysis; the programmer does not supply an explicit passport for this.

Looking at the table in fig \ref{synch-syntax}, the first thing that requires CAL is the input interface where variants and patterns are used. The use of a variant on a channel in any transition on that channel immediately suggests the top level of the term structure associated with it. For example, a channel $c$ that is tested on variants
\verb$?v$ and \verb$?w$ in transitions has a term comparable with
\begin{verbatim}
(: v:$vdata, w:$wdata || $rest :)
\end{verbatim}
where the three variables are fresh and represent the terms for the variants \verb$v$, \verb$w$ and the choice term that contains the rest of the variants, respectively.  In the absence of any other usage of $c$ in the synchroniser program, the above choice term for $c$ will be the one included in the passport. Now suppose that there is a transition in the code on \verb$?v(x,y||z)$, that is, variant \verb$v$ is known to contain two integers \verb$x$ and \verb$y$ and the rest of the message \verb"$z". Consequently the following constraint will be included in the passport
\begin{verbatim}
$vdata = <p:{x:int, y:int || $z}) (not p): [int,int || $z]>
);
\end{verbatim}
Here \verb"p" is a fresh flag and \verb"$z" a fresh term variable. This way the synchroniser can be matched with a producer that produces records containing integer fields \verb$x$ and \verb$y$ or one that produces parameter lists where integer values \verb$x$ and \verb$y$ are the first and second members of the list, respectively. In both cases the structure of the rest of the message, be it a list or a record, can be learned by the synchroniser by examining the term assigned to the MDL variable \verb"$z".

Now consider the output interface. Suppose channel $b$ is used in a send clause as follows:
\verb"send (g,w,k)=>b". Assume that  variables \verb$g$ and \verb$w$ are store, channel or tail variables,
and that \verb"k" is a state variable or an alias of an integer. The term for channel $b$ will then be as follows
\begin{verbatim}
(
	union $g $w < p:{k:int}, (not p):[int] >
)
\end{verbatim}
where the term variables \verb"$g" and \verb"$k" are justaposed with the terms associated with store (or similar) variables \verb"g" and \verb"w" in the synchroniser program.
\end{document}